\documentclass [11pt]{IEEEtran}
\usepackage{graphicx}
\usepackage{amssymb}
\usepackage{epsfig}
\usepackage{amsmath}

\usepackage{doublespace}
\newcommand{\thbs}{\mbox{\small \boldmath{$\theta$}}}
\newcommand{\deft}{{\stackrel{\triangle}{=}}}
\newcommand{\eg}{{\em e.g., }}

\newcommand{\diag}{\operatorname{diag} \,}

\newcommand{\ml}{\operatorname{ML}}

\newcommand{\tr}{\operatorname{Tr}}

\newcommand{\bbi}{{{\bf I}}}
\newcommand{\bbh}{{{\bf H}}}
\newcommand{\bbu}{{{\bf U}}}
\newcommand{\bbq}{{{\bf Q}}}

\newcommand{\bbd}{{{\bf D}}}

\newcommand{\bbx}{{{\bf x}}}
\newcommand{\bby}{{{\bf y}}}
\newcommand{\bbp}{{{\bf P}}}
\newcommand{\bh}{{{\bf h}}}

\newcommand{\bx}{{\bbx}}

\newcommand{\by}{{\bby}}
\newcommand{\bw}{{\bf w}}

\newcommand{\A}{{\mathcal{A}}}

\newcommand{\R}{{\mathcal{R}}}

\newcommand{\RR}{{\mathbb{R}}}

\newcommand{\bl}{\left(}
\newcommand{\br}{\right)}
\newcommand{\blc}{\left\{}
\newcommand{\brc}{\right\}}

\newcommand{\bba}{{\mathbf A}}
\newcommand{\bbl}{{\mathbf L}}

\newcommand{\bbc}{{\mathbf C}}

\newcommand{\bbv}{{\mathbf V}}
\newcommand{\bu}{{\mathbf u}}
\newcommand{\bv}{{\mathbf v}}

\newcommand{\thb}{\mbox{\boldmath{$\theta$}}}
\newcommand{\hthb}{\hat{\mbox{\boldmath{$\theta$}}}}

\newtheorem{theorem}{Theorem}

\newtheorem{proposition}{Proposition}

\title{\singlespace Generalized SURE for Exponential Families: Applications to Regularization}
\author{Yonina C. Eldar\thanks{Department of Electrical Engineering,
Technion---Israel Institute of Technology, Haifa 32000, Israel.
Phone: +972-4-8293256, fax: +972-4-8295757, E-mail:
yonina@ee.technion.ac.il. This work was supported in part by the
Israel Science Foundation under Grant no. 1081/07 and by the
European Commission in the framework of the FP7 Network of
Excellence in Wireless COMmunications NEWCOM++ (contract no.
216715).}}
\date{\today}

\begin{document}
\maketitle

\begin{abstract}
Stein's unbiased risk estimate (SURE) was proposed by Stein for
the independent, identically distributed (iid) Gaussian model in
order to derive estimates that dominate least-squares (LS). In
recent years, the SURE criterion has been employed in a variety of
denoising problems for choosing regularization parameters that
minimize an estimate of the mean-squared error (MSE). However, its
use has been limited to the iid case which precludes many
important applications. In this paper we begin by deriving a SURE
counterpart for general, not necessarily iid distributions from
the exponential family. This enables extending the SURE design
technique to a much broader class of problems. Based on this
generalization we suggest a new method for choosing regularization
parameters in penalized LS estimators. We then demonstrate its
superior performance over the conventional generalized cross
validation approach and the discrepancy method in the context of
image deblurring and deconvolution. The SURE technique can also be
used to design estimates without predefining their structure.
However, allowing for too many free parameters impairs the
performance of the resulting estimates. To address this inherent
tradeoff we propose a regularized SURE objective. Based on this
design criterion, we derive a wavelet denoising strategy that is
similar in sprit to the standard soft-threshold approach but can
lead to improved MSE performance.
\end{abstract}

\section{Introduction}
Estimation in multivariate problems is a fundamental topic in
statistical signal processing. One of the most common recovery
strategies for deterministic unknown parameters is the well-known
maximum likelihood (ML) method. The ML estimator enjoys several
appealing properties, including asymptotic efficiency under
suitable regularity conditions. Nonetheless, its mean-squared
error (MSE) can be improved upon in the non-asymptotic regime in
many different settings.

In their seminal work, Stein and James showed that for the
independent, identically-distributed (iid) linear Gaussian model,
it is possible to construct a nonlinear estimator with lower MSE
than that of ML for all values of the unknowns \cite{S56,JS61}.
Various modifications of the James-Stein method have since been
developed that are applicable to the non-iid Gaussian case as well
\cite{S71,A73,B76,BE05,BE05j,EBN032}.

The James-Stein approach is based on the Stein unbiased risk
estimate (SURE) \cite{s73,S81}, which is an unbiased estimate of
the MSE. Since the MSE in general depends on the true unknown
parameter values it cannot be used as a design objective. However,
using the SURE principle leads to a relatively simple technique
for determining methods that have lower MSE than ML. The idea is
to choose a class of estimates, and then select the member from
the class that minimizes the SURE estimate of the MSE. This
strategy has been applied to a variety of different denoising
techniques \cite{dj95,zd98,lbu07,bp05}. Typically, in these
problems, implicit prior information on the signal to be recovered
is incorporated into the chosen structure of the estimate. For
example, in wavelet denoising the signal is assumed to be sparse
in the wavelet domain which motives the use of thresholding. Only
the value of the threshold is determined by the SURE principle.

The SURE method is appealing as it allows to directly approximate
the MSE of an estimate from the data, without requiring knowledge
of the true parameter values. However, it has two main drawbacks
which severely limit its use in practical applications. The first
restriction is that it was originally limited to the iid Gaussian
case. Several extensions have been developed for different
independent models. In particular, a SURE principle for iid,
infinitely divisible random variables with finite variance is
derived in \cite{AH06}. Extensions to independent variables from a
continuous exponential family are treated in \cite{H78,B80}, while
the discrete exponential case is discussed in \cite{H82s}. All of
these generalizations are confined to the independent case which
precludes a variety of important applications such as image
deblurring.

The second drawback of using SURE as a design criterion is that in
order to get meaningful estimators the basic structure of the
estimate must be determined in advance. If no parametrization is
assumed, then there are too many free variables to be optimized,
and the SURE method will typically not lead to good MSE behavior.

In this paper we extend the basic SURE principle in two
directions, in order to circumvent the two fundamental drawbacks
outlined above. First, we generalize SURE to multivariate,
possibly non-iid exponential families. In particular, we develop
an unbiased estimate of the MSE for a general Gaussian vector
model.
 Exponential probability density functions (pdfs) play an important
role in statistics due to the Pitman-Koopman-Darmois theorem
\cite{P36,D35,K36}, which states that among distributions whose
domain does not vary with the parameter being estimated, only in
exponential families is there a sufficient statistic with bounded
dimension as the sample size increases \cite{LC98}. Furthermore,
efficient estimators exist only when the underlying model is
exponential. Many known distributions are of the exponential form,
such as Gaussian, gamma, chi-square, beta, Dirichlet, Bernoulli,
binomial, multinomial, Poisson, and geometric distributions. Our
result has important practical value as it extends the
applicability of the SURE technique to more general estimation
models, and in particular to scenarios in which the observations
are dependent. This is the situation, for example, when using
overcomplete wavelet transforms, and in image deblurring. In
addition, we derive results for the case in which the model is
rank-deficient so that the pdf depends only on a projection of the
parameter vector.

An immediate application of this extension is to the general
linear Gaussian model. In this setup, we seek to estimate a
parameter vector $\thb$ from noisy, blurred measurements
$\bx=\bbh\thb+\bw$ where $\bw$ is a Gaussian noise vector and
$\bbh$ may be rank deficient. One of the most popular recovery
strategies in this context is the regularized least-squares (LS)
method. In this approach, the estimate is designed to minimize a
regularized LS objective where typical choices of penalization are
weighted $\ell_2$ or $\ell_1$ norms. An important aspect of this
technique, which significantly impacts its performance, is
selecting the regularization parameter. A variety of different
methods have been proposed for this purpose
\cite{R86e,DG95,GK92,H93,HH93,B96,MKM99,K05}. When an $\ell_2$
norm is used to penalize the LS solution, the resulting estimate
is linear and is referred to as Tikhonov regularization. One of
the most popular regularization selection methods in this context
is generalized cross-validation (GCV) \cite{ghw79}. When an
$\ell_1$ penalty is used, the resulting estimate is nonlinear so
that applying the GCV approach is more complicated. An alternative
choice is the discrepancy method in which the parameter is chosen
such that the resulting data error is equal to the noise variance.

Here, we suggest an alternative strategy based on our extended
SURE criterion. Specifically, we use SURE to evaluate the MSE of
the penalized solution for any choice of regularization, and then
select the value that minimizes the SURE estimate. This allows
SURE-based optimization of a broad class of deblurring and
deconvolution methods including both linear and nonlinear
techniques. A similar approach was studied in the special case of
Tikhonov regularization with white noise in \cite{R86e,DG95,GK92}.
However, our technique is not limited to an $\ell_2$ penalty and
can be used with any other penalized LS method.  When the estimate
is not given explicitly but rather as a solution of an
optimization problem we can still employ the SURE strategy by
using a Monte-Carlo approach to approximate the derivative of the
estimate, which figures in the SURE expression \cite{RBU08}. Using
several test images and a deconvolution problem, we demonstrate
that this strategy often leads to significant performance
improvement over the standard GCV and discrepancy selection
criteria in the context of image deblurring and deconvolution.

Finally, to circumvent the need for pre-defining a particular
structure when applying SURE, we propose an alternative approach
based on regularizing the SURE objective. Specifically, we suggest
adding a penalization term to the SURE expression and choosing an
estimate that minimizes the regularized function. In this way, we
can control the properties of the estimate without having to
a-priori assume a specific structure. We then illustrate this
strategy in the context of wavelet denoising. Instead of assuming
a threshold estimate and choosing the threshold to minimize the
SURE criterion, as in \cite{dj95}, we design an estimate that
minimizes an $\ell_1$ regularized SURE objective. The resulting
denoising scheme has the form of a threshold with a particular
form of shrinkage, that is different than that obtained when using
soft or hard thresholding. To evaluate our method, we compare it
with SureShrink of Donoho and Johnstone \cite{dj95}, by repeating
the simulations reported in their paper. As we show, the recovery
results tend to be better using our technique. Moreover, our
approach is general as it is not tailored to a specific problem.
We thus believe that using a regularized SURE principle together
with the generalized SURE developed here can extend the
applicability of SURE-based estimators to a broad class of
problems.

The remaining of the paper is organized as follows. In
Section~\ref{sec:mse} we introduce the basic concept of MSE
estimation. An extension of SURE to exponential families is
developed in Section~\ref{sec:exp}. In Section~\ref{sec:rd} we
discuss rank-deficient models in which the pdf of the data depends
only on a projection of the unknown parameter vector. We then
specialize the results to the linear Gaussian model in
Section~\ref{sec:gauss}. Applications to regularization selection
are discussed in Section~\ref{sec:reg}. The regularized SURE
criterion, together with an application to wavelet denoising, are
developed in Section~\ref{sec:rsure}.

\section{MSE Estimation}
\label{sec:mse}

We denote vectors by boldface lowercase letters, \eg $\bx$, and
matrices by boldface uppercase letters \eg $\bba$. The $i$th
component of a vector $\by$ is written as $y_i$, and
$\hat{(\cdot)}$ is an estimated vector. The identity matrix is
written as $\bbi$, $\bba^T$ is the transpose of $\bba$, and
$\bba^\dagger$ denotes the pseudo-inverse. For a length-$m$ vector
function $\bh(\bu) \in \RR^m$ of $\bu \in \RR^m$, \begin{equation}
\tr\bl \frac{\partial \bh(\bu)}{\partial \bu}
\br=\sum_{i=1}^m\frac{\partial h_i(\bu)}{\partial u_i}.
\end{equation}

We consider the class of problems in which our goal is to estimate
a deterministic parameter vector $\thb$ from observations $\bx$
which are related through a pdf $f(\bx;\thb)$. We further assume
that the pdf belongs to the exponential family of distributions
and can be expressed in the form
\begin{equation}
\label{eq:model} f(\bx;\thb)=r(\bx)\exp\{\thb^T
\phi(\bx)-g(\thb)\},
\end{equation}
where $r(\bx)$ and $\phi(\bx)$ are functions of the data only, and
$g(\thb)$ depends on the unknown parameter $\thb$.

As an example of an application where the model (\ref{eq:model})
can occur, consider the location problem of estimating a parameter
vector $\thb \in \RR^m$ from observations $\bx \in \RR^n$ related
through the linear model:
\begin{equation}
\label{eq:lg} \bx=\bbh\thb+\bw,
\end{equation}
where $\bw$ is a zero-mean Gaussian random vector with covariance
$\bbc \succ 0$. The pdf of $\bx$ is then given by (\ref{eq:model})
with
\begin{eqnarray}
\label{eq:gauss} & r(\bx)=\frac{1}{\sqrt{(2\pi)^n
\det(\bbc)}}\exp\{-(1/2)\bx^T\bbc^{-1}\bx\}; & \nonumber \\
& \phi(\bx)=\bbh^T\bbc^{-1}\bx; & \nonumber \\
& g(\thb)=(1/2)\thb^T\bbh^T\bbc^{-1}\bbh\thb. &
\end{eqnarray}
Other examples of distributions in the exponential family include
Poisson with unknown mean, exponential with unknown mean, gamma,
and Bernoulli or binomial with unknown success probabilities.

Given the model (\ref{eq:model}), a sufficient statistic for
estimating $\thb$ is given by
\begin{equation}
\label{eq:ss} \bu=\phi(\bx).
\end{equation}
Therefore, any reasonable estimate of $\bx$ will be a function
only of $\bu$. More specifically, from the Rao-Blackwell theorem
\cite{K93} it follows that if $\hat{\thb}$ is an estimate of
$\thb$ which is not a function only of $\bu$, then the estimate
$E\{ \hthb|\bu\}$ has lesser or equal MSE than that of $\hthb$,
for all $\thb$. Therefore, in the sequel, we only consider methods
that depend on the data via $\bu$.

Let $\hthb$ be an arbitrary estimate of $\thb$, which we would
like to design to minimize the MSE, defined by
$E\{\|\hthb-\thb\|^2\}$. In practice, $\hthb=\bh(\bu)$ where
$\bh(\bu)$ is some function of $\bu$ that is typically chosen to
have a particular structure, parameterized by a vector
$\mbox{\boldmath{$\alpha$}}$. For example, $\bh(\bu)=\alpha \bu$
where $\alpha$ is a scalar, or $h_i(\bu)=\psi_\alpha(u_i)$ where
\begin{equation}
\psi_{\alpha}(u)=\mbox{sign}(u)[|u|-\alpha]_+
\end{equation}
 is a
soft-threshold with cut-off $\alpha$. Ideally, we would like to
select $\alpha$ to minimize the MSE. Since this is impossible, as
we show below, instead in the SURE approach
$\mbox{\boldmath{$\alpha$}}$ is designed to minimize an unbiased
estimate (referred to as the SURE estimate) of the MSE.

We can express the MSE of $\hthb=\bh(\bu)$ as
\begin{eqnarray}
\label{eq:mse} E\blc \|\hthb-\thb\|^2\brc =
\|\thb\|^2+E\blc\|\bh(\bu)\|^2\brc-2E\blc \bh^T(\bu) \thb\brc.
\end{eqnarray}
In order to minimize the MSE over $\bh(\bu)$ we need to explicitly
evaluate the expression
\begin{equation}
\label{eq:f} v(\bh,\thb)=E\blc\|\bh(\bu)\|^2\brc-2E\blc \bh^T(\bu)
\thb\brc.
\end{equation}
Evidently, the MSE will depend in general on $\thb$, which is
unknown, and therefore cannot be minimized. Instead, we may seek
an unbiased estimate of $v(\bh,\thb)$ and then choose $\bh$ to
minimize this estimate. The difficult expression to approximate is
$E\{ \bh^T(\bu) \thb\}$ as the dependency on $\thb$ is explicit.
Therefore, we concentrate on estimating this term. If this can be
done, then we can easily obtain an unbiased MSE estimate.
Specifically, suppose we construct a function $g(\bh(\bu))$ that
depends only on $\bu$ (and not on $\thb$), such that
\begin{equation}
E \blc g(\bh(\bu))\brc =E\blc \bh^T(\bu) \thb\brc \deft
E_{\bh,\thbs}.
\end{equation}
Then
\begin{equation}
\label{eq:sp} \hat{v}(\bh)=\|\bh(\bu)\|^2-2g(\bh(\bu)),
\end{equation}
is an unbiased estimate of $v(\bh,\thb)$, since clearly
$E\{\hat{v}(\bh)\}=v(\bh,\thb)$. A reasonable strategy therefore
is to select $\bh(\bu)$ to minimize our assessment $\hat{v}(\bh)$
of the MSE. This approach was first proposed by Stein in
\cite{s73,S81} for the iid Gaussian model (\ref{eq:lg}) with
$\bbc=\bbi$ and $\bbh=\bbi$.

The design framework proposed above reduces to finding an unbiased
estimate of $E_{\bh,\thbs}$. Clearly, any such approximation will
depend on the pdf $f(\bx;\thb)$. In the next section we develop an
unbiased estimate when the pdf is given by the exponential model
(\ref{eq:model}). Before addressing the general setting, to ease
the exposition we illustrate the main idea proposed by Stein, by
first considering the simpler iid Gaussian case. In  this setting
we seek to estimate a vector $\thb \in \RR^m$ from measurements
$\bx=\thb+\bw$, where $\bw$ is a zero-mean Gaussian random vector
with iid components of variance $\sigma^2$. In
Section~\ref{sec:rd} we treat the more difficult case in which
$\bu$ lies in a subspace $\A$ of $\RR^m$, and the pdf
(\ref{eq:model}) depends on $\thb$ only through its orthogonal
projection onto $\A$. This situation arises, for example, in the
linear Gaussian model (\ref{eq:lg}) when $\bbh$ is rank deficient.
For this setup, we develop a SURE estimate of the MSE in
estimating the projected parameter. In Sections~\ref{sec:gauss}
and \ref{sec:reg} we consider several examples of estimates in
which the free parameters are chosen to minimize the SURE
objective. In particular, we propose alternatives to the popular
GCV and discrepancy methods for regularization. In
Section~\ref{sec:rsure}, we suggest a regularized SURE strategy
for determining $\bh(\bu)$ without the need for parametrization,
and demonstrate its performance in the context of wavelet
denoising.

\newpage
\section{Extended SURE Principle}
\label{sec:exp}

\subsection{IID Gaussian Model}

We begin our development by treating the iid Gaussian setting.
Since from (\ref{eq:gauss}), $\bu=(1/\sigma^2)\bx$, we consider
estimates $\hthb=\bh(\bx)$ that are a function of $\bx$.

To develop an unbiased estimate of $E_{\bh,\thbs}$, we exploit the
fact that for the Gaussian pdf $f(\bx;\thb)$
\begin{equation}
\label{eq:ip} (x_i-\theta_i) f(\bx;\thb)=-\sigma^2 \frac{\partial
f(\bx;\thb)}{\partial x_i}.
\end{equation}
Assuming that $E\{|h_i(\bx)|\}$ is bounded and $h_i(\bx)$ is
weakly differentiable\footnote{Roughly speaking, a function is
weakly differentiable if it has a derivative almost everywhere, as
long as the points that are not differentiable are not delta
functions; see \cite{LL01} for a more formal definition.} in
$\bx$, we have that
\begin{eqnarray}
\label{eq:int2s} E_{\bh,\thbs}& = & \sum_{i=1}^m
\int_{-\infty}^\infty h_i(\bx)\theta_i f(\bx;\thb)d\bx \nonumber
\\
& = & \sum_{i=1}^m \int_{-\infty}^\infty h_i(\bx)\bl x_i
f(\bx;\thb)+\sigma^2 \frac{\partial f(\bx;\thb)}{\partial x_i}\br
d\bx \nonumber
\\
& = & E\{\bh^T(\bx)\bx\}+\sigma^2\sum_{i=1}^m
\int_{-\infty}^\infty h_i( \bx)\frac{\partial
f(\bx;\thb)}{\partial x_i}d\bx,
\end{eqnarray}
where the second equality is a result of (\ref{eq:ip}). To
evaluate the second term in (\ref{eq:int2s}), we use integration
by parts:
\begin{equation}
\label{eq:int2s2} \int_{-\infty}^\infty h_i( \bx)\frac{\partial
f(\bx;\thb)}{\partial x_i}d\bx=-\int_{-\infty}^\infty h_i'(
\bx)f_i(\bx;\thb)d\bx=-E\{h_i'(\bx)\},
\end{equation}
were we denoted $h_i'(\bx)=\partial h_i(\bx)/\partial x_i$, and
used the fact that $|h_i(\bx)f(\bx;\thb)| \rightarrow 0$ for
$|x_i| \rightarrow \infty$ since $E\{|h_i(\bx)|\}$ is bounded. We
conclude from (\ref{eq:int2s}) and (\ref{eq:int2s2}) that
\begin{equation}
E_{\bh,\thbs}= - \sigma^2\sum_{i=1}^m
E\{h'_i(\bx)\}+E\{\bh^T(\bx)\bx\},
\end{equation}
and therefore,
\begin{equation}
\label{eq:setg}  - \sigma^2\sum_{i=1}^m \frac{\partial
h_i(\bx)}{\partial x_i}+\bh^T(\bx)\bx
\end{equation}
 is an unbiased estimate of $E_{\bh,\thbs}$.
\subsection{Extended SURE}

We now extend the basic approach outlined in the previous section
to the general class of exponential pdfs. In order to address this
model, we only consider methods that depend on the data via $\bu$.
This enables the use of integration by parts, similar to the iid
Gaussian setting.

The following theorem provides an unbiased estimate of $E\blc
\bh^T(\bu) \thb\brc$ which depends only on $\bu$ and not on the
unknown parameters $\thb$.
\begin{theorem}
\label{thm:sure} Let $\bx$ denote a random vector with exponential
pdf given by (\ref{eq:model}), and let $\bu=\phi(\bx)$  be a
sufficient statistic for estimating $\thb$ from $\bx$. Let
$\bh(\bu)$ be an arbitrary function of $\thb$ that is weakly
differentiable in $\bu$ and such that $E \blc |h_i(\bu)| \brc$ is
bounded. Then
\begin{equation}
\label{eq:thm} E\blc \bh^T(\bu) \thb\brc= -E\blc\tr\bl
\frac{\partial \bh(\bu)}{\partial \bu}\br\brc -E\blc
\bh^T(\bu)\frac{\partial\ln q(\bu)}{\partial\bu}\brc,
\end{equation}
where
\begin{equation}
\label{eq:qu} q(\bu)= \int r(\bx) \delta(\bu-\phi(\bx))d\bx,
\end{equation}
and $\delta(\bx)$ is the Kronecker delta function.
\end{theorem}
Note, that as we show in the proof of the theorem, the pdf
$f(\bu;\thb)$ of $\bu$ is given by
\begin{equation}
\label{eq:qua} f(\bu;\thb)= q(\bu)\exp\{\thb^T \bu-g(\thb)\}.
\end{equation}
Therefore, an alternative to computing $q(\bu)$ using
(\ref{eq:qu}) is to evaluate the pdf of $\bu$ and then use
(\ref{eq:qua}).

From the theorem, it follows that
\begin{equation}
\label{eq:set} -\tr\bl \frac{\partial \bh(\bu)}{\partial \bu}\br -
\bh^T(\bu)\frac{\partial\ln q(\bu)}{\partial\bu}\,
\end{equation}
is an unbiased estimate of $E\{ \bh^T(\bu) \thb\}$. In the iid
Gaussian case, $\bu=(1/\sigma^2) \bx$ so that
\begin{equation}
\label{eq:lg1} \frac{\partial \bh(\bu)}{\partial \bu}=\sigma^2
\frac{\partial \bh(\bx)}{\partial \bx}.
\end{equation}
Furthermore, $\bu$ is a Gaussian iid vector with elements that
have mean $(1/\sigma^2) \thb$ and variance $1/\sigma^2$ so that
$q(\bu)$, which is the normalization factor, is given by $
q(\bu)=K\exp\{-\|\bu\|^2\sigma^2/2\}$ for a constant $K$.
Consequently
\begin{equation}
\label{eq:lg2} \frac{\partial \ln q(\bu)}{\partial
\bu}=-\sigma^2\bu=-\bx.
\end{equation}
Substituting (\ref{eq:lg1}) and (\ref{eq:lg2}) into
(\ref{eq:set}), the estimate reduces to (\ref{eq:setg}), derived
in the iid Gaussian setting.

\begin{proof}
To prove the theorem we first determine the pdf of $\bu$. Since
$\bu=\phi(\bx)$ we have that \cite[p. 127]{K93}
\begin{equation}
\label{eq:pu} f(\bu;\thb)=\int
f(\bx;\thb)\delta(\bu-\phi(\bx))d\bx.
\end{equation}
Substituting (\ref{eq:model}) into (\ref{eq:pu}),
\begin{eqnarray}
f(\bu;\thb)  & = & \exp\{\thb^T \bu-g(\thb)\} \int r(\bx)
\delta(\bu-\phi(\bx))d\bx \nonumber \\
& = & q(\bu)\exp\{\thb^T \bu-g(\thb)\}.
\end{eqnarray}
Now,
\begin{eqnarray}
\label{eq:int} \lefteqn{E\blc  \bh^T(\bu) \thb \brc =} \nonumber \\
& = &\int \bh^T(\bu) \thb \exp\{\thb^T
\bu-g(\thb)\} q(\bu) d \bu \nonumber \\
& = & \sum_{i=1}^m \int  h_i(\bu) \theta_i \exp\{\thb^T
\bu-g(\thb)\} q(\bu) d \bu.
\end{eqnarray}
Noting that
\begin{equation}
 \theta_i \exp\{\thb^T \bu-g(\thb)\}=\frac{\partial \exp\{\thb^T
\bu-g(\thb)\}}{\partial u_i},
\end{equation}
we have
\begin{eqnarray}
\label{eq:int2} \lefteqn{\int_{-\infty}^\infty  h_i(\bu) \theta_i
\exp\{\thb^T \bu-g(\thb)\} q(\bu) du_i=} \nonumber \\ & = &
\int_{-\infty}^\infty h_i(\bu) q(\bu) \frac{\partial \exp\{\thb^T
\bu-g(\thb)\}}{\partial u_i} du_i \nonumber \\
& = &  - \int_{-\infty}^\infty \frac{\partial
h_i(\bu)q(\bu)}{\partial u_i}\exp\{\thb^T \bu-g(\thb)\}
  du_i,
\end{eqnarray}
where we used the fact that $|h_i(\bu) q(\bu) \exp\{\thb^T
\bu-g(\thb)\}| \rightarrow 0$ for $|u_i| \rightarrow \infty$ since
$E\{|h_i(\bu)| \}$ is bounded. Now,
\begin{equation}
\label{eq:intt} \frac{\partial h_i(\bu)q(\bu)}{\partial
u_i}=\frac{\partial h_i(\bu)}{\partial u_i}q(\bu)+\frac{\partial
q(\bu)}{\partial u_i}h_i(\bu).
\end{equation}
Substituting (\ref{eq:int2}) and (\ref{eq:intt}) into
(\ref{eq:int}),
\begin{eqnarray}
\label{eq:int3}\lefteqn{ E\blc  \bh^T(\bu) \thb \brc=} \nonumber \\
& = & -\sum_{i=1}^m \int \frac{\partial h_i(\bu)q(\bu)}{\partial
u_i}\exp\{\thb^T
\bu-g(\thb)\}\partial \bu \nonumber \\
& = & \sum_{i=1}^m \bl-E\blc \frac{\partial h_i(\bu)}{\partial
u_i}\brc-E\blc \frac{\partial q(\bu)}{\partial
u_i}\frac{h_i(\bu)}{q(\bu)} \brc \br\nonumber \\
& = &-E\blc \tr\bl \frac{\partial \bh(\bu)}{\partial
\bu}\br\brc-E\blc \bh^T(\bu)\frac{\partial\ln q(\bu)}{\partial
\bu}\brc,
\end{eqnarray}
which completes the proof of the theorem.
\end{proof}

Based on Theorem~\ref{thm:sure} we can develop a generalized SURE
principle for estimating an unknown parameter vector $\thb$ in an
exponential model. Specifically, let $\hthb=\bh(\bu)$ be an
arbitrary estimate of $\thb$ based on the data $\bx$, where
$\bh(\bu)$ satisfies the regularity conditions of
Theorem~\ref{thm:sure}. Combining (\ref{eq:mse}) and
Theorem~\ref{thm:sure}, an unbiased estimate of the MSE of $\hthb$
is given by
\begin{equation}
\label{eq:spf} S(\bh)=\|\thb\|^2+\|\bh(\bu)\|^2 +2 \tr\bl
\frac{\partial \bh(\bu)}{\partial \bu}\br +
2\bh^T(\bu)\frac{\partial\ln q(\bu)}{\partial \bu}.
\end{equation}
We may then design $\hthb$ by choosing $\bh(\bu)$ to minimize
$S(\bh)$.

In the iid Gaussian case, (\ref{eq:spf}) reduces to
\begin{equation}
\label{eq:sures} \|\thb\|^2+\|\bh(\bx)\|^2+ 2\sigma^2
\frac{\partial \bh(\bx)}{\partial \bx}-2\bh^T(\bx)\bx
\end{equation}
where we used the fact that $\bu=(1/\sigma^2)\bx$ which implies
$\partial\ln q(\bu)/\partial \bu=-\bx$, and $\partial
\bh(\bu)/\partial \bu=\sigma^2 \partial \bh(\bx)/\partial \bx$.
The MSE estimate (\ref{eq:sures}) was first proposed by
Stein\footnote{We note that the expression obtained by Stein is
slightly different since instead of optimizing $\hthb=\bh(\bx)$,
he considered estimates of the form $\hthb=\bx+\bh(\bx)$ and then
optimized $\bh(\bx)$. The two expressions differ by a constant,
which does not effect the optimization of $\bh(\bx)$.} in
\cite{s73,S81}.

The SURE estimate can be used to determine unknown regularization
parameters which comprise a given estimation strategy. An example
is the SureShrink approach to wavelet denoising \cite{dj95}.
Extending this technique, our general SURE objective
(\ref{eq:spf}) may be used to select regularization parameters in
more general models. We discuss these ideas in the context of
linear Gaussian problems in Section~\ref{sec:gauss}.

\section{Rank-Deficient Models}
\label{sec:rd}

In some settings, the sufficient statistic $\bu$ lies in a
subspace $\A$ of $\RR^m$. As an example, suppose that in the
Gaussian model (\ref{eq:lg}) $\bbh$ is rank-deficient. In this
case $\bu=\bbh^T \bbc^{-1}\bx$ lies in the range space
$\R(\bbh^T)$ of $\bbh^T$, which is a subspace of $\RR^m$. If
$\thb$ is not restricted to a subspace, then we do not expect to
be able to reliably estimate $\thb$ from $\bu$, unless some
additional information on $\thb$ is known. Nonetheless, we may
still obtain a reliable assessment of the part of $\thb$ that lies
in $\A$. Denote by $\bbp$ the orthogonal projection onto $\A$.
Then, we show below, that if $\bu$ depends on $\thb$ only through
$\bbp \thb$, and in addition $\bu$ has an exponential pdf, then we
can obtain a SURE estimate of the error in $\A$, namely $E\{\|\bbp
\hthb-\bbp \thb\|^2\}$. If in addition $\hthb$ lies in $\A$, then
up to a constant, independent of $\hthb$, this approximation is
also an unbiased estimate of the true MSE $E\{\|
\hthb-\thb\|^2\}$.

To derive the SURE estimate in this case, we first note that if
$\bu$ lies in $\A$, then
\begin{equation}
\thb^T\bu=(\bbp\thb)^T(\bbp\bu).
\end{equation}
Suppose that $\A$ has dimension $r<m$. Since $\bbp \thb$ lies in
an $r$-dimensional space, it can be expressed in terms of $r$
components in an appropriate basis. Denoting by $\bbv$ an $m
\times r$ matrix with orthonormal vectors that span $\A=\R(\bbp)$,
the vector $\bbp\thb$ can be expressed as $\bbp\thb=\bbv\thb'$ for
an appropriate length-$r$ vector $\thb'$. Similarly,
$\bbp\bu=\bbv\bu'$. Therefore, we can write
\begin{equation}
\thb^T\bu=\thb'^T\bu',
\end{equation}
where we used the fact that $\bbv^T\bbv=\bbi$.  We assume that
$\bu'$ is a sufficient statistic for $\thb'$ and that
$f(\bu';\thb')$ has an exponential pdf:
\begin{equation}
\label{eq:pdfp} f(\bu';\thb')=q(\bu')\exp\{\thb'^T
\bu'-g(\thb')\}.
\end{equation}
Under this assumption, we next derive a SURE estimate of the MSE
in $\A$.

The MSE in estimating $\thb$ can then be written as
\begin{equation}
E\{\|\hthb-\thb\|^2\}=E\{\|\bbp \hthb-\bbp
\thb\|^2\}+E\{\|(\bbi-\bbp)\hthb-(\bbi-\bbp)\thb\|^2\}.
\end{equation}
If $\hthb$ lies in $\A$, then $(\bbi-\bbp)\hthb=0$ and the second
term is constant, independent of $\hthb$. Therefore, in this case,
to optimize $\hthb$ it is sufficient to obtain an unbiased
estimate of the first term.
 As we show below, such an assessment can be derived using
similar ideas to those in Theorem~\ref{thm:sure}. Even if $\hthb$
does not lie in $\A$, the SURE estimate we develop may be used to
approximate the first term. Since $\bu$ depends only on
$\bbp\thb$, it is reasonable to restrict attention to estimates
$\hthb =\bh_{\mbox{\boldmath{$\alpha$}}}(\bu)$ where the
parameters $\mbox{\boldmath{$\alpha$}}$ are tuned to minimize the
MSE in $\A$ $E\{\|\bbp \hthb-\bbp \thb\|^2\}$, subject to any
other prior information we may have, such as norm constraints on
$\thb$. In such cases we can use a regularized SURE criterion with
the SURE objective being the MSE in $\A$, as we discuss in
Section~\ref{sec:rsure}.

We now develop a SURE estimate of the MSE $E\{\|\bbp \hthb-\bbp
\thb\|^2\}$. To this end we need to find an unbiased estimate of
\begin{equation}
E \{ \hthb^T\bbp \thb\}= E \{ \hthb^T\bbv\thb' \}=E \{
(\bbv^T\hthb)^T\thb' \}
\end{equation}
Let $\hthb=\bh(\bu)=\bh(\bu')$ (since $\bu=\bbv\bu'$, it is clear
that $\hthb$ is a function of $\bu'$). By our assumption, $\thb'$
has an exponential pdf with sufficient statistic $\bu'$.
Therefore, we can apply Theorem~\ref{thm:sure} to $\bbv^T
\bh(\bu)$ for any function $\bh(\bu)$ that obeys the conditions of
the theorem, which yields
\begin{eqnarray}
\label{eq:thmp} E\blc \bh^T(\bu)\bbv \thb' \brc & = &
-E\blc\tr\bl \bbv^T\frac{\partial
 \bh(\bu')}{\partial \bu'}\br\brc -E\blc \bh^T(\bu')\bbv \frac{\partial\ln
q(\bu')}{\partial\bu'}\brc \nonumber \\
 & = &  -E\blc\tr\bl \bbp\frac{\partial
 \bh(\bu)}{\partial \bu}\br\brc -E\blc \bh^T(\bu)\bbv \frac{\partial\ln
q(\bu')}{\partial\bu'}\brc,
\end{eqnarray}
where we used the fact that $\bbp=\bbv\bbv^T$ and
\begin{equation}
\frac{\partial
 \bh(\bu')}{\partial \bu'}=\frac{\partial
 \bh(\bu)}{\partial \bu} \bbv.
\end{equation}

We conclude that if $\bh(\bu)$ is an estimate of a parameter
vector $\thb$, where $\bu$ lies in a subspace $\A$ and is a
sufficient statistic for estimating $\bbp \thb$, with $\bbp$
denoting the orthogonal projection onto $\A$, then an unbiased
estimate of the MSE $E\{\|\bbp \bh(\bu)-\bbp \thb\|^2\}$ is given
by
\begin{equation}
\label{eq:psure} S(\bh)=\|\bbp \thb\|^2+\|\bbp \bh(\bu)\|^2+2
\tr\bl \bbp\frac{\partial
 \bh(\bu)}{\partial \bu}\br +2 \bh^T(\bu)\bbv \frac{\partial\ln
q(\bu')}{\partial\bu'},
\end{equation}
with $\bu=\bbv\bu'$ and $\bbv$ denoting an orthonormal basis for
$\A$. When $\A=\RR^m$, $\bbp=\bbi, \bbv=\bbi$ and (\ref{eq:psure})
reduces to (\ref{eq:spf}).

\section{Linear Gaussian Model}
\label{sec:gauss}

We now specialize the SURE principle to the linear Gaussian model
(\ref{eq:lg}). We begin by treating the case in which $\bbh$ is an
$n \times m$ matrix with $n \geq m$ and full column rank. We then
discuss the rank-deficient scenario.

\subsection{Full-Rank Model}

To use Theorem~\ref{thm:sure} we need to compute the pdf $q(\bu)$
of $\bu$. Since $\bu=\bbh^T\bbc^{-1}\bx$, it is a Gaussian random
vector with mean $\bbh^T\bbc^{-1}\bbh\thb$ and covariance
$\bbh^T\bbc^{-1}\bbh$. As $q(\bu)$ is the function multiplying the
exponential in the pdf of $\bu$, it follows from (\ref{eq:gauss})
that
\begin{equation}
\label{eq:qlgf}  q(\bu)= K \exp\{-(1/2)\bu^T
(\bbh^T\bbc^{-1}\bbh)^{-1}\bu \},
\end{equation}
where $K$ is a constant, independent of $\bu$. Therefore,
\begin{equation}
\label{eq:dqlg} \frac{\partial\ln q(\bu)}{\partial
\bu}=-(\bbh^T\bbc^{-1}\bbh)^{-1}\bu=-\hthb_{\ml},
\end{equation}
where $\hthb_{\ml}$ is the ML estimate of $\thb$ given by
\begin{equation}
\label{eq:mlg}
\hthb_{\ml}=(\bbh^T\bbc^{-1}\bbh)^{-1}\bbh^T\bbc^{-1}\bx.
\end{equation}
 It then follows from Theorem~\ref{thm:sure} that
\begin{equation}
\label{eq:thmg} E\blc \bh^T(\bu) \thb \brc= -E \blc \tr\bl
\frac{\partial \bh(\bu)}{d \bu}\br-\bh^T(\bu)\hthb_{\ml}\brc.
\end{equation}
Using (\ref{eq:spf}) and (\ref{eq:dqlg}) we conclude that
\begin{equation}
S(\bh)=\|\thb\|^2+\|\bh(\bu)\|^2+2 \bl\tr\bl \frac{\partial
\bh(\bu)}{\partial \bu}\br-\bh^T(\bu)\hthb_{\ml}\br,
\end{equation}
is an unbiased estimate of the MSE.

\subsection{Rank-Deficient Model}

Next, we consider the linear Gaussian model (\ref{eq:lg}) with a
rank-deficient $\bbh$.

Suppose that $\bbh$ has a singular value decomposition
$\bbh=\bbu\Sigma \bbq^T$ for some unitary matrices $\bbu$ and
$\bbq$. Let $\bbh$ have rank $r$ so that $\Sigma$ is a diagonal $n
\times m$ matrix what the first $r$ diagonal elements equal to
$\sigma_i>0$ and the remaining elements equal $0$. In this case,
$\bbv$ is equal to the first $r$ columns of $\bbq$ and
$\thb'=\bbv^T \thb$. A sufficient statistic for estimating $\thb'$
is $\bu'=\bbv^T\bbh^T\bbc^{-1}\bx$. Indeed, $\bu'$ is a Gaussian
random vector with mean $\mu'=\bbv^T\bbh^T\bbc^{-1}\bbh\thb$ and
covariance $\bbc'=\bbv^T\bbh^T\bbc^{-1}\bbh\bbv$. Using the SVD of
$\bbh$ we have that
\begin{eqnarray}
\mu' & = &  \Lambda[\bbu^T\bbc^{-1}\bbu]_r\thb', \nonumber \\
 \bbc' & = & \Lambda[\bbu^T\bbc^{-1}\bbu]_r,
\end{eqnarray}
where $\Lambda$ is an $r \times r$ diagonal matrix with diagonal
elements $\sigma_i^2>0$ and $[\bba]_r$ is the $r \times r$
top-left principle block of size $r$ of the matrix $\bba$. Since
$\bbc \succ 0$, $\bbc'$ is invertible. Therefore,
\begin{equation}
f(\bu';\thb')=q(\bu') \exp\{\thb'\bu'-g(\thb') \}
\end{equation}
with
\begin{eqnarray}
\label{eq:gaussp} & q(\bu')=\frac{1}{\sqrt{(2\pi)^n
\det(\bbc')}}\exp\{-(1/2)\bu'^T\bbc'^{-1}\bu'\}; & \nonumber \\
& g(\thb')=(1/2)\thb'^T\Lambda[\bbu^T\bbc^{-1}\bbu]_r\thb'. &
\end{eqnarray}

We therefore conclude from (\ref{eq:psure}) that an unbiased
estimate of the MSE $E\{\|\bbp \bh(\bu)-\bbp\thb\|^2 \}$ is given
by
\begin{eqnarray}
\|\bbp\thb\|^2+\|\bbp\bh(\bu)\|^2+2\bl\tr\bl \bbp \frac{\partial
 \bh(\bu)}{\partial \bu}\br-\bh^T(\bu) \hthb_{\ml}\br,
\end{eqnarray}
where
$\hthb_{\ml}=(\bbh^T\bbc^{-1}\bbh)^\dagger\bbh^T\bbc^{-1}\bx$ is
an ML estimate. Here we used the fact that \sloppy
$\bbv\bbc'^{-1}\bu'=(\bbh^T\bbc^{-1}\bbh)^\dagger\bbh^T\bbc^{-1}\bx$.

We summarize our results for the linear Gaussian model in the
following proposition.
\begin{proposition}
\label{prop:sureg} Let $\bx$ denote measurements of an unknown
parameter vector $\thb$ in the linear Gaussian model
(\ref{eq:lg}), where $\bw$ is a zero-mean Gaussian random vector
with covariance $\bbc \succ 0$. Let $\bh(\bu)$ with
$\bu=\bbh^T\bbc^{-1}\bx$ be an arbitrary function of $\thb$ that
is weakly differentiable in $\bu$ and such that $E \blc |h_i(\bu)|
\brc$ is bounded, and let $\bbp$ be an orthogonal projection onto
$\R(\bbh^T)$. Then
\[
 E\blc \bh^T(\bu)\bbp \thb \brc= -E \blc
\tr\bl \bbp \frac{\partial \bh(\bu)}{\partial
\bu}\br-\bh^T(\bu)\hthb_{\ml}\brc,
\]
where
\[\hthb_{\ml}=(\bbh^T\bbc^{-1}\bbh)^\dagger\bbh^T\bbc^{-1}\bx\] is an ML estimate of $\thb$.
 An unbiased estimate of the MSE
$E\{\|\bbp \bh(\bu)-\bbp \thb\|^2\}$
 is
\begin{equation}
\label{eq:spg} S(\bh)=\|\bbp\thb\|^2+\|\bbp\bh(\bu)\|^2+2
\bl\tr\bl \bbp\frac{\partial \bh(\bu)}{\partial
\bu}\br-\bh^T(\bu)\hthb_{\ml}\br.
\end{equation}
\end{proposition}

\subsection{Examples}

To illustrate the use of the SURE principle, suppose that we
consider estimators of the form $\hthb=\alpha \hthb_{\ml}$ where
$\hthb_{\ml}$ is given by (\ref{eq:mlg}), and we would like to
select a good choice of $\alpha$. To this end, we minimize the
SURE unbiased estimate of the MSE given by
Proposition~\ref{prop:sureg} with $\bh(\bu)=\alpha\hthb_{\ml}$.
Note that in this case $\bh(\bu) \in \R(\bbh^T)$  so that
$S(\bh)+\|(\bbi-\bbp)\thb\|^2$ is an unbiased estimate of the
total MSE $E\{\|\hthb-\thb\|^2\}$ and therefore it suffices to
minimize $S(\bh)$.

For this choice of $\bh(\bu)$, minimizing $S(\bh)$ is equivalent
to minimizing
\begin{equation}
\alpha^2\|\hthb_{\ml}\|^2+2\bl \alpha\tr \bl
(\bbh^T\bbc^{-1}\bbh)^\dagger\br-\alpha \|\hthb_{\ml}\|^2\br.
\end{equation}
The optimal choice of $\alpha$ is
\begin{equation}
\alpha=1- \frac{\tr \bl
(\bbh^T\bbc^{-1}\bbh)^\dagger\br}{\|\hthb_{\ml}\|^2},
\end{equation}
resulting in
\begin{equation}
\label{eq:jsm} \hthb=\bl 1- \frac{\tr \bl
(\bbh^T\bbc^{-1}\bbh)^\dagger\br}{\|\hthb_{\ml}\|^2}\br
\hthb_{\ml}.
\end{equation}
The estimate of (\ref{eq:jsm}) coincides with the balanced blind
minimax method proposed in \cite[Eq. (45)]{BE05j}, which was
derived based on a minimax framework \cite{EBN032}. Here we see
that the same technique results from applying our generalized SURE
criterion. A striking feature of this estimate, proved in
\cite{BE05j}, is that when $\bbh^T\bbc^{-1}\bbh$ is invertible and
its effective dimension is larger than $4$, it dominates ML for
all values of $\thb$ (see Theorem 3 in \cite{BE05j}). This means
that its MSE is always lower than that of the ML method,
regardless of the true value of $\thb$.

When $\bbh=\bbi$ and $\bbc=\sigma^2\bbi$,  (\ref{eq:jsm}) reduces
to
\begin{equation}
\label{eq:jss} \hthb=\bl 1- \frac{n\sigma^2}{\|\bx\|^2}\br \bx,
\end{equation}
which coincides with Stein's estimate \cite{S56}. This technique
is known to dominate ML for $n \geq 3$.

If in addition we require that $\alpha \geq 0$, then the estimate
of (\ref{eq:jsm}) becomes
\begin{equation}
\label{eq:jsmp} \hthb=\left[ 1- \frac{\tr \bl
(\bbh^T\bbc^{-1}\bbh)^\dagger\br}{\|\hthb_{\ml}\|^2}\right]_+
\hthb_{\ml},
\end{equation}
where we used the notation
\begin{equation}
[x]_+=\left\{
\begin{array}{ll}
x, & x \geq 0;\\
 0, & x < 0.
\end{array}
 \right.
\end{equation}
The method of (\ref{eq:jsmp}) is a positive-part version of
(\ref{eq:jsm}). In the iid case, it reduces to the positive-part
Stein's estimate \cite{B64}, which is known to dominate the
standard Stein approach (\ref{eq:jss}).

 Next, consider the case in which
$\bbh=\bbi$ and $\bbc=\bbd$ with
$\bbd=\diag(\sigma_1^2,\ldots,\sigma_n^2)$ and suppose we seek a
diagonal estimate of the form $\hat{\theta}_i=\alpha_i x_i$.
Minimizing the unbiased estimate of (\ref{eq:spg}) in this case is
equivalent to minimizing
\begin{equation}
\sum_{i=1}^n \alpha_i^2x_i^2+2\sum_{i=1}^n
\sigma_i^2\alpha_i-2\sum_{i=1}^n\alpha_i x_i^2,
\end{equation}
which yields
\begin{equation}
\label{eq:scs} \alpha_i= 1- \frac{\sigma_i^2}{x_i^2}.
\end{equation}
Restricting the coefficients $\alpha_i$ to be non-negative leads
to the estimate
\begin{equation}
\label{eq:jsc} \hat{\theta}_i= \left[1-
\frac{\sigma_i^2}{x_i^2}\right]_+x_i.
\end{equation}

In contrast to $\hthb$ of (\ref{eq:jsm}), which dominates the ML
method, it can be proven that the estimate of (\ref{eq:jsc}) is
not dominating. Thus, we see that by allowing for too many free
parameters, we impair the performance of the SURE-based estimate.
On the other hand, assuming strong structure, as in
(\ref{eq:jsm}), severely restricts the class of estimators and
consequently limits the possible performance advantage which can
be obtained. In Section~\ref{sec:rsure} we suggest a regularized
SURE strategy in order to overcome this inherent tradeoff between
over-parametrization and performance.

\section{Application to Regularization Selection}
\label{sec:reg}

A popular strategy for solving inverse problems of the form
(\ref{eq:lg}) is to use regularization techniques in conjunction
with a LS objective. Specifically, the estimate $\hthb$ is chosen
to minimize a regularized LS criterion:
\begin{equation}
\label{eq:reg} (\bx-\bbh\hthb)\bbc^{-1}(\bx-\bbh\hthb)+\lambda
\|\bbl \hthb\|
\end{equation}
where the norm is arbitrary. Here $\bbl$ is some regularization
operator such as the discretization of a first or second order
differential operator that accounts for smoothness properties of
$\thb$, and $\lambda$ is the regularization parameter
\cite{HH93,H93}.  An important problem in practice is the
selection of $\lambda$, which strongly effects the recovery
performance. One of the most popular approaches to choosing
$\lambda$ when the estimate is linear (as is the case when a
squared-$\ell_2$ norm is used in (\ref{eq:reg})) is the
generalized cross-validation (GCV) method \cite{ghw79}. When the
estimate takes on a more complicated nonlinear form, a popular
selection method is the discrepancy principle \cite{GK92}.

Based on our generalized SURE criterion, we choose $\lambda$ to
minimize the SURE objective (\ref{eq:spg}). As we demonstrate for
the cases in which the norm in (\ref{eq:reg}) is the
squared-$\ell_2$ or $\ell_1$ norms, this method can dramatically
outperform GCV and the discrepancy technique in practical
applications.

\subsection{Image Deblurring}

We first consider the case in which the squared-$\ell_2$ norm is
used in (\ref{eq:reg}). The solution then has the form
\begin{equation}
\label{eq:tik} \hthb=(\bbq+\lambda
\bbl^T\bbl)^{-1}\bbh^T\bbc^{-1}\bx,
\end{equation}
where for brevity we denoted
\begin{equation}
\bbq=\bbh^T\bbc^{-1}\bbh.
\end{equation}
 The estimate (\ref{eq:tik}) is
commonly referred to as Tikhonov regularization \cite{TA77}.

In the GCV method, $\lambda$ is chosen to minimize
\begin{equation}
\label{eq:gcv} G(\lambda)=\frac{1}{\tr^2(\bbi-(\bbq+\lambda
\bbl^T\bbl)^{-1}\bbq)}\sum_{i=1}^n (\bx_i-[\bbh \hthb]_i)^2.
\end{equation}
The SURE choice follows directly from minimizing (\ref{eq:spg}).
In our simulations below, both minimizations where performed by
using the \texttt{fmincon} function on Matlab.

\begin{figure}[h]
\begin{center}
\resizebox{!}{2
in}{\includegraphics{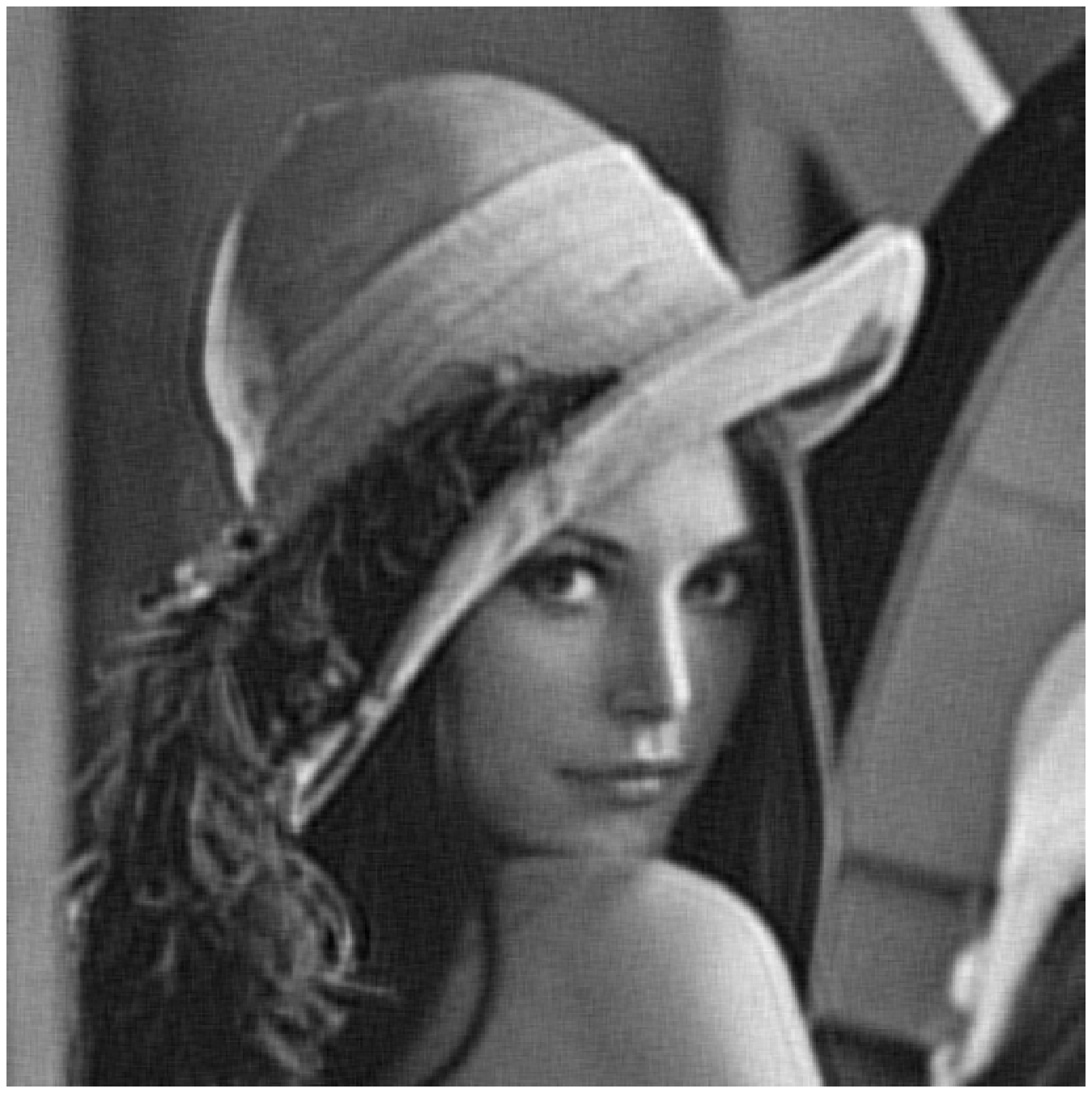}}\resizebox{!}{2
in}{\includegraphics{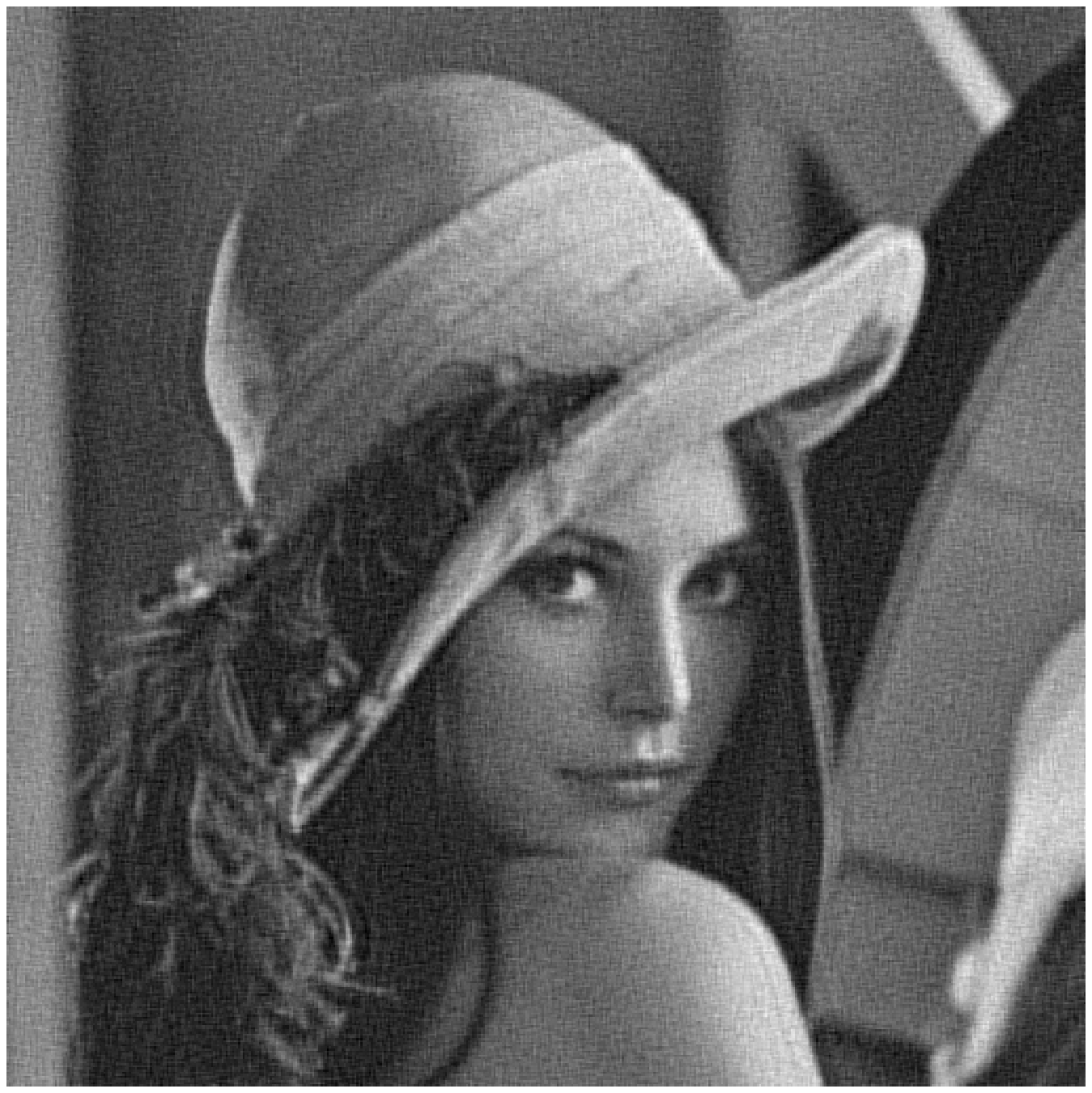}}

 (a) \hspace*{1.9in} (b)

\resizebox{!}{2
in}{\includegraphics{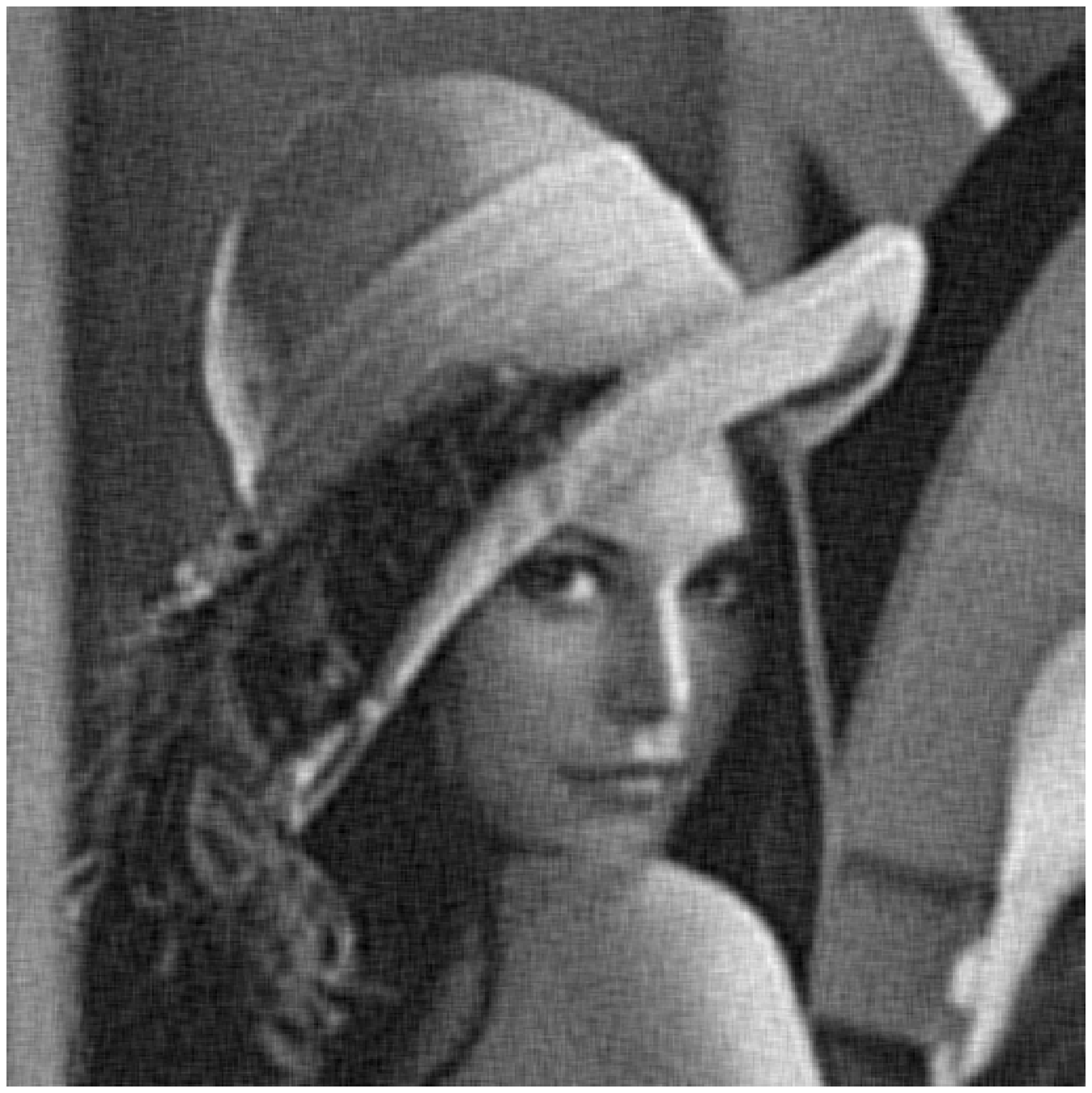}}\resizebox{!}{2
in}{\includegraphics{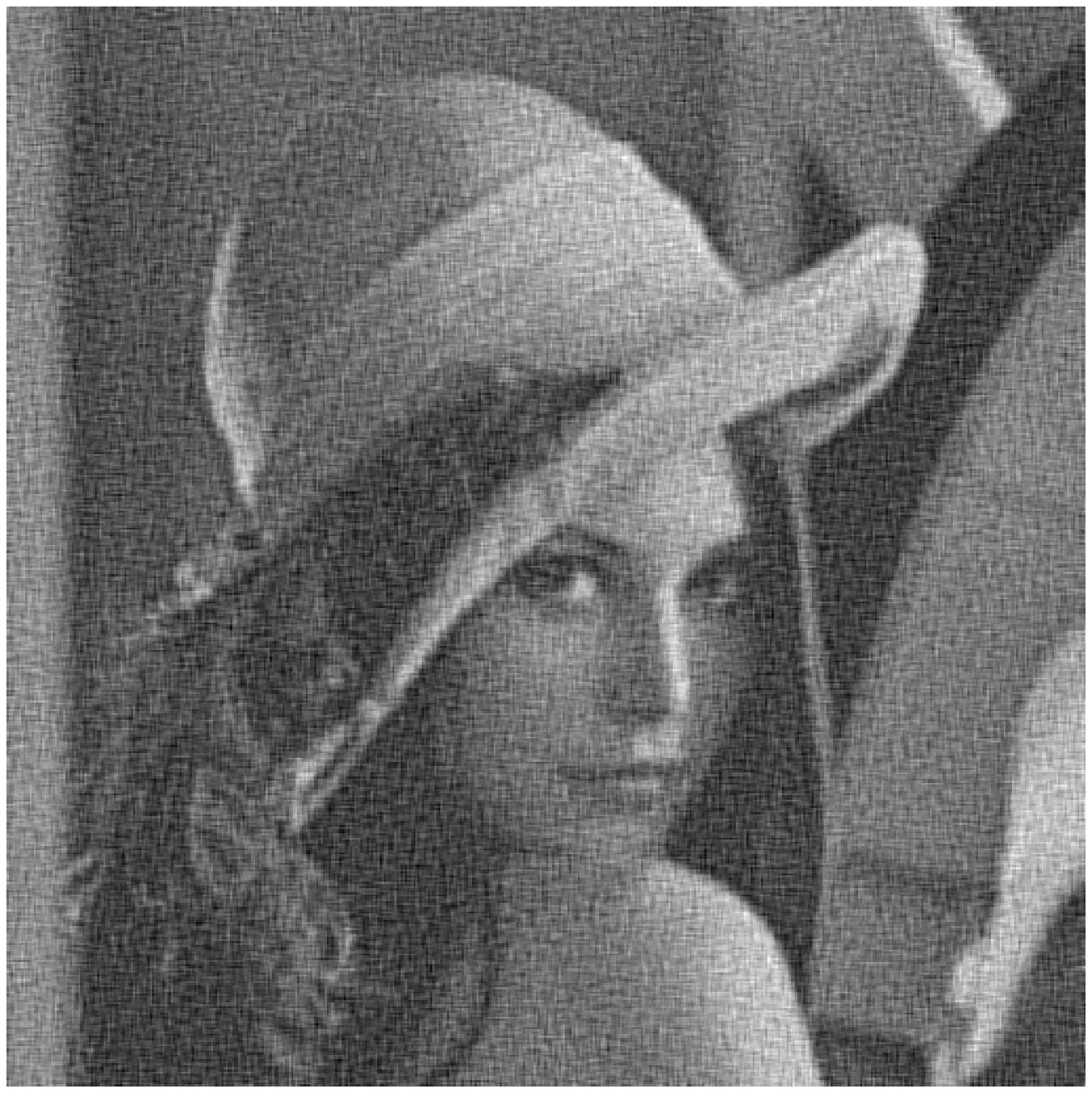}}

 (c) \hspace*{1.9in} (d)

\resizebox{!}{2
in}{\includegraphics{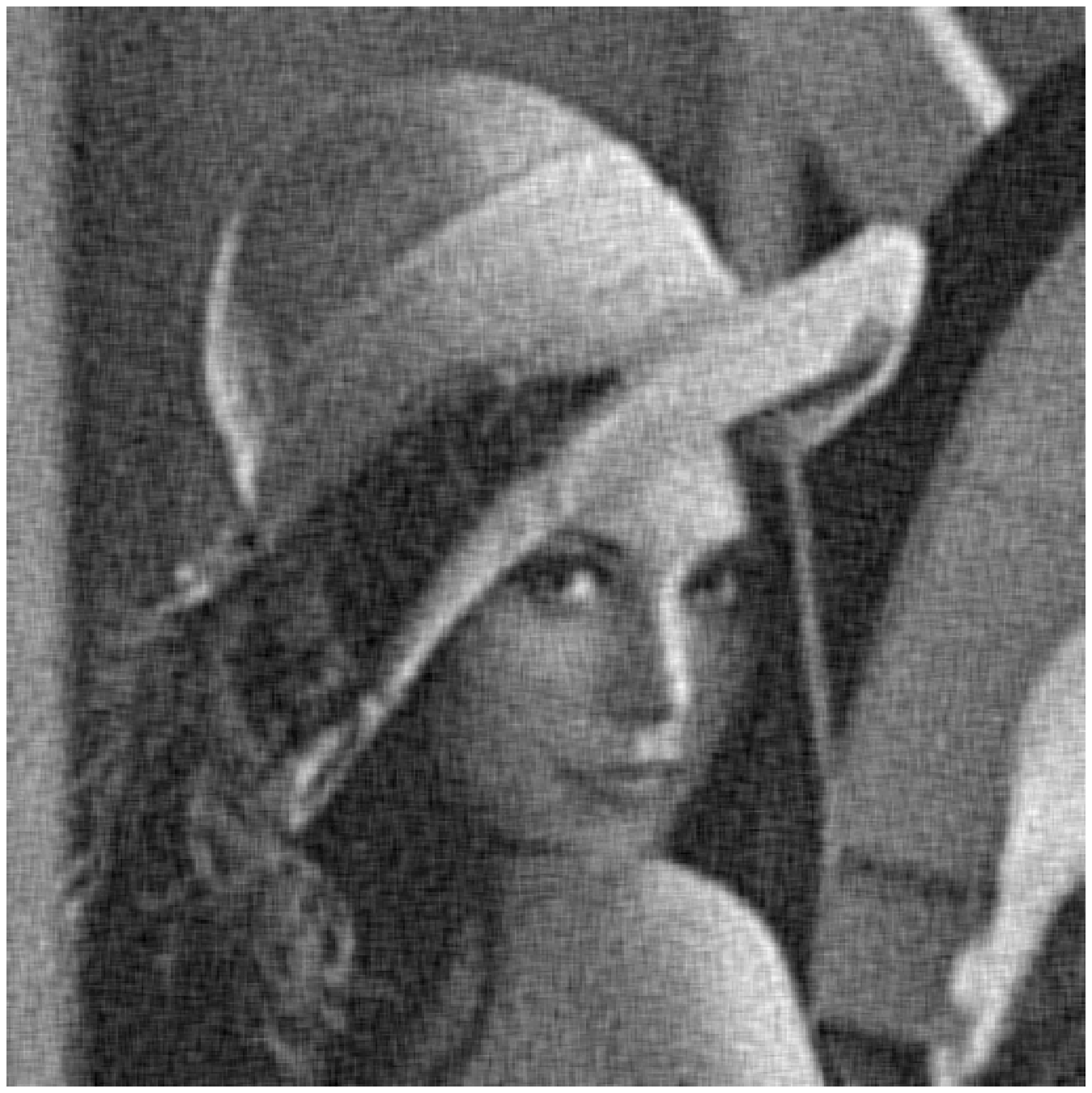}}\resizebox{!}{2
in}{\includegraphics{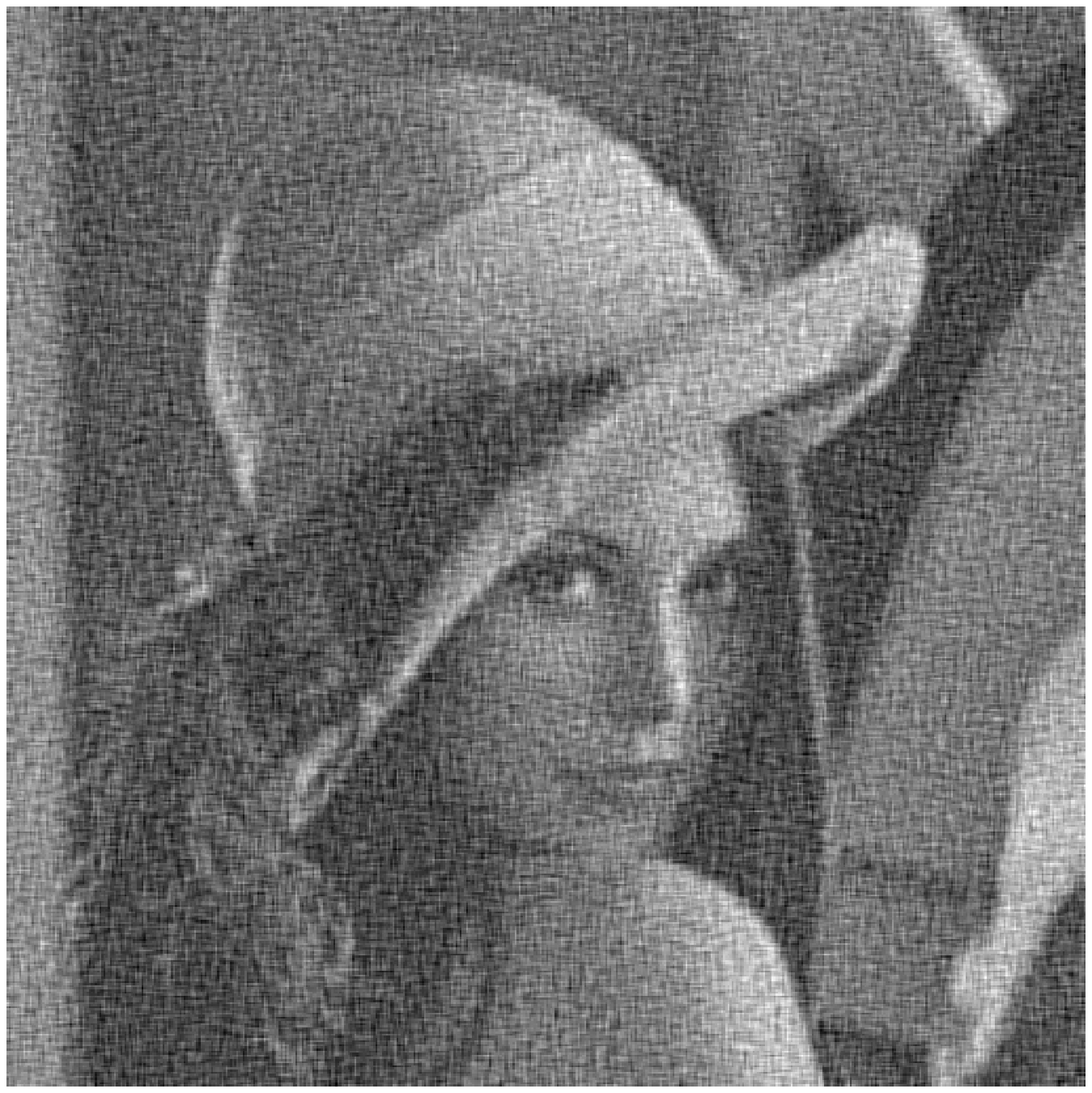}}

 (e) \hspace*{1.9in} (f)

\caption{Deblurring of Lena using Tikhonov regularization with
SURE (left) and GCV (right) choices of regularization and
different noise levels: (a), (b) $\sigma=0.01$ (c),(d)
$\sigma=0.05$ (e),(f) $\sigma=0.1$. }
 \label{fig:lena}
\end{center}
\end{figure}

\begin{figure}[h]
\begin{center}
\resizebox{!}{2
in}{\includegraphics{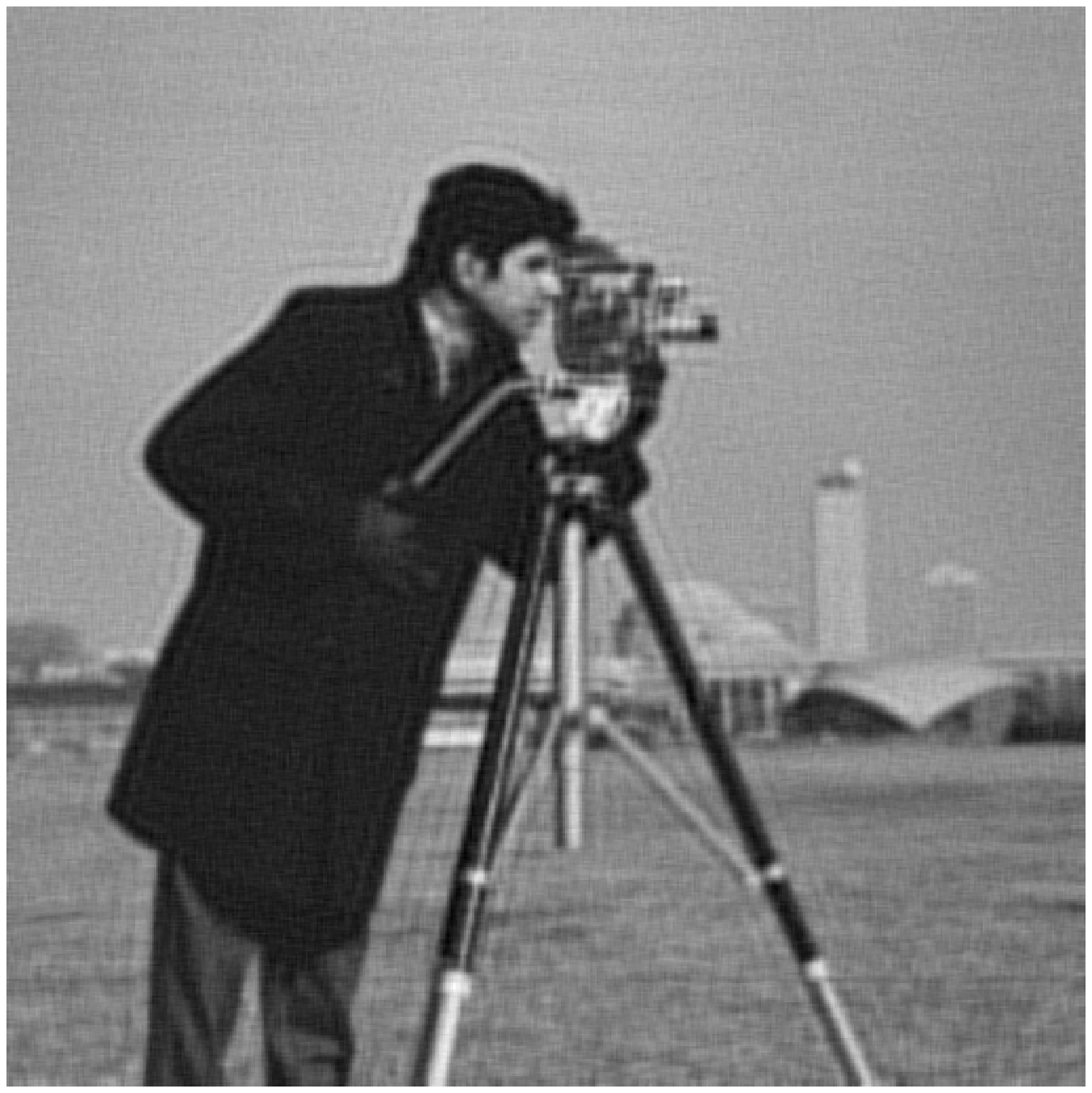}}\resizebox{!}{2
in}{\includegraphics{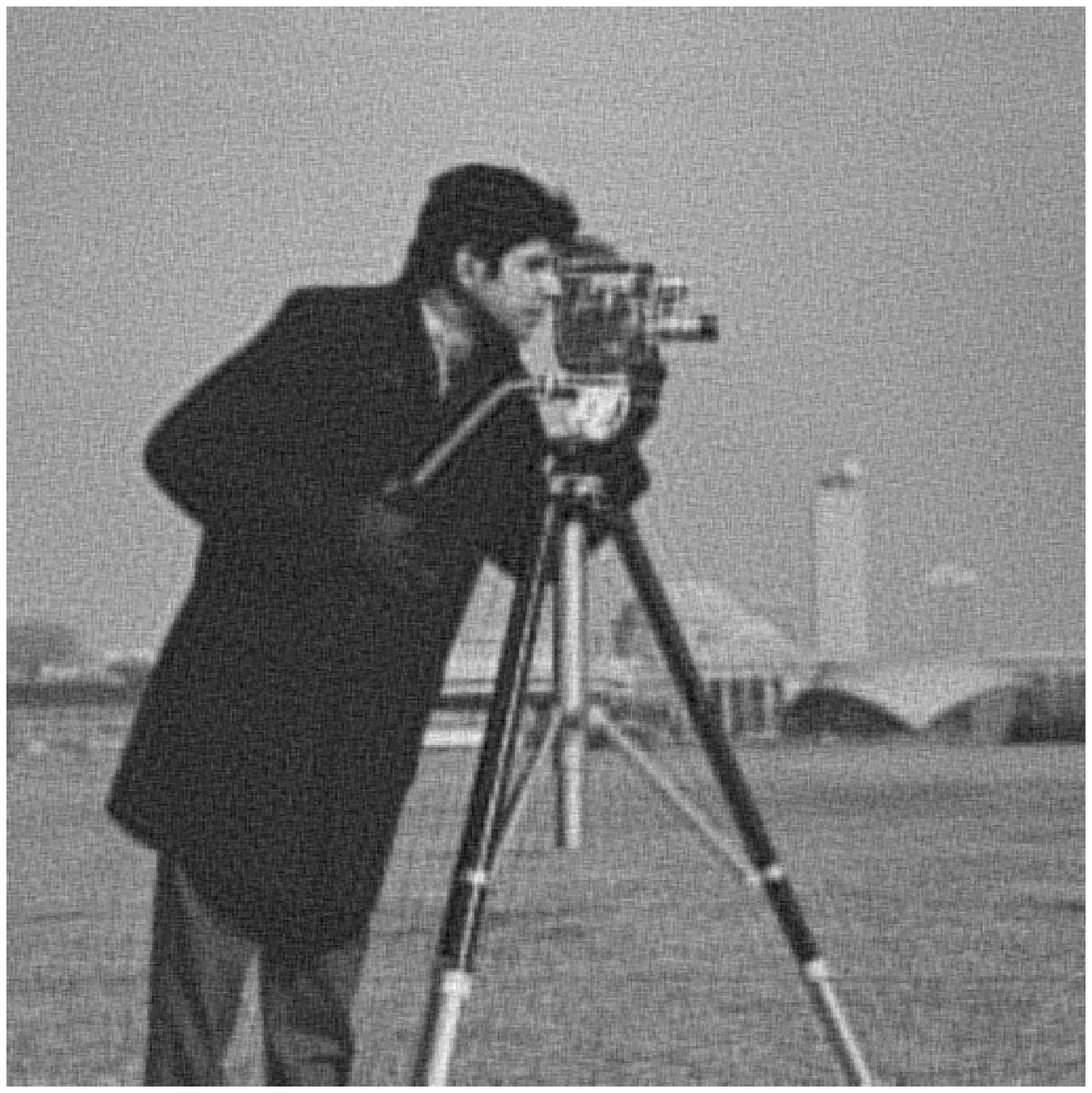}}

 (a) \hspace*{1.9in} (b)

\resizebox{!}{2
in}{\includegraphics{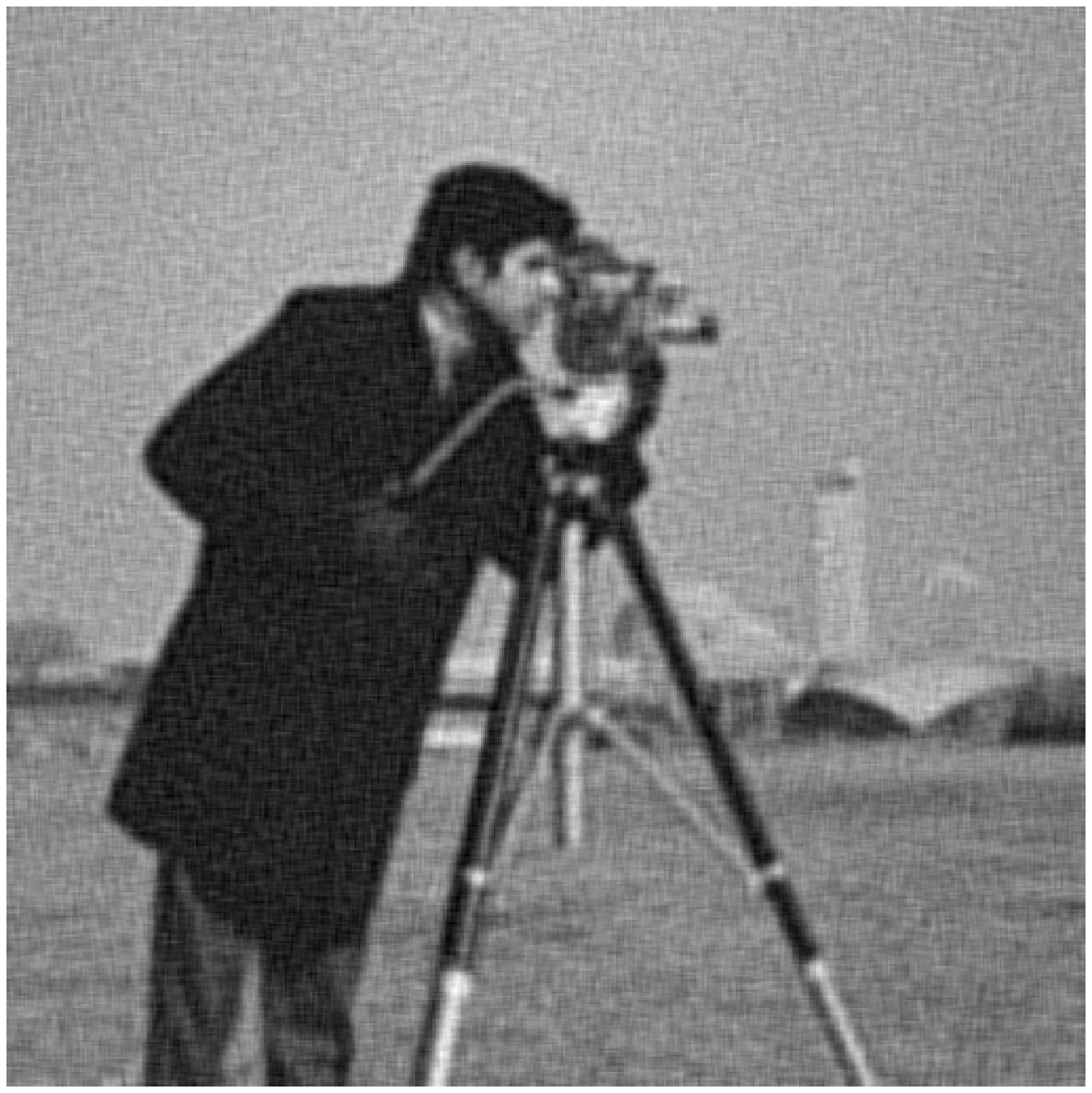}}\resizebox{!}{2
in}{\includegraphics{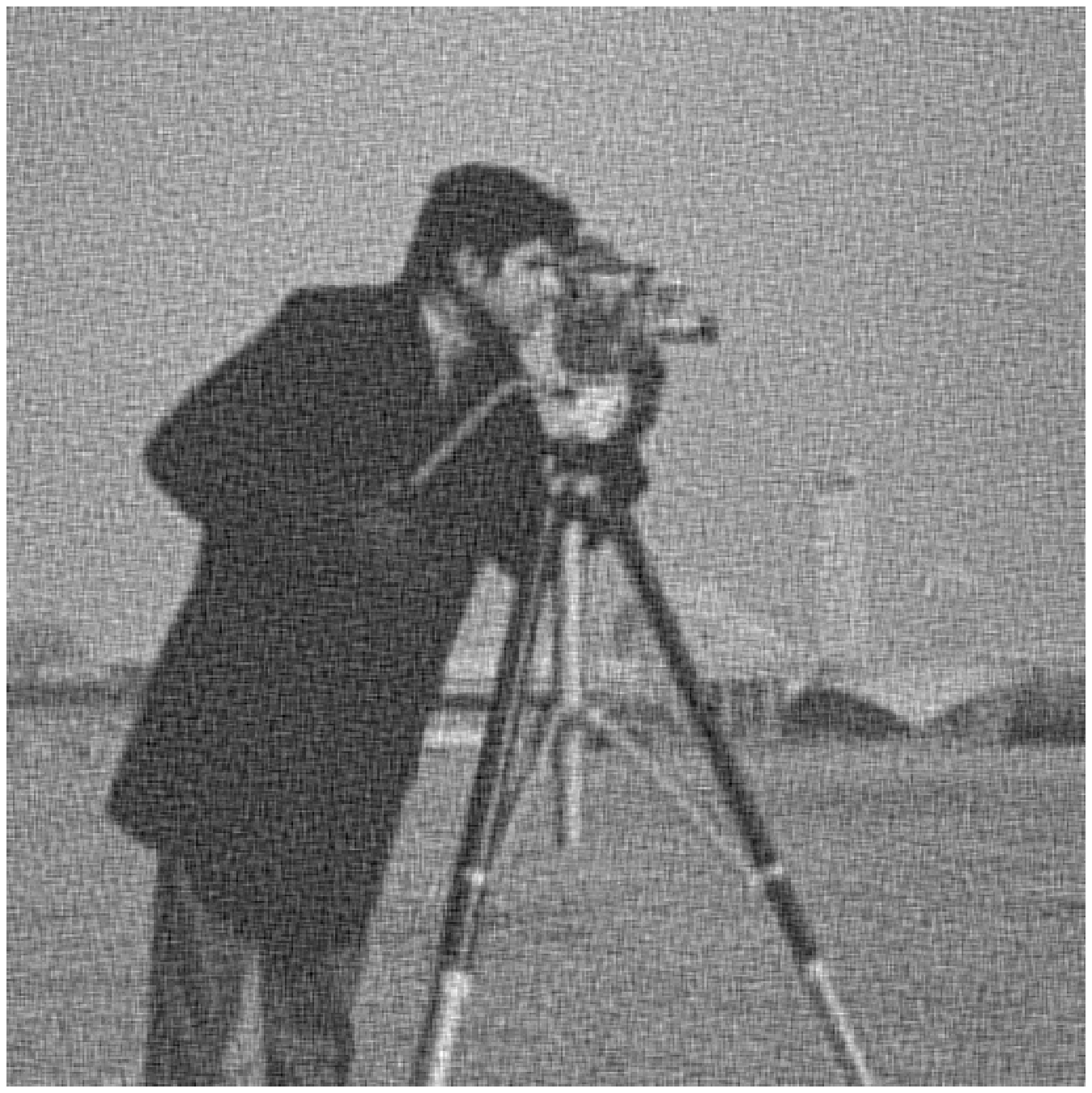}}

 (c) \hspace*{1.9in} (d)

\resizebox{!}{2
in}{\includegraphics{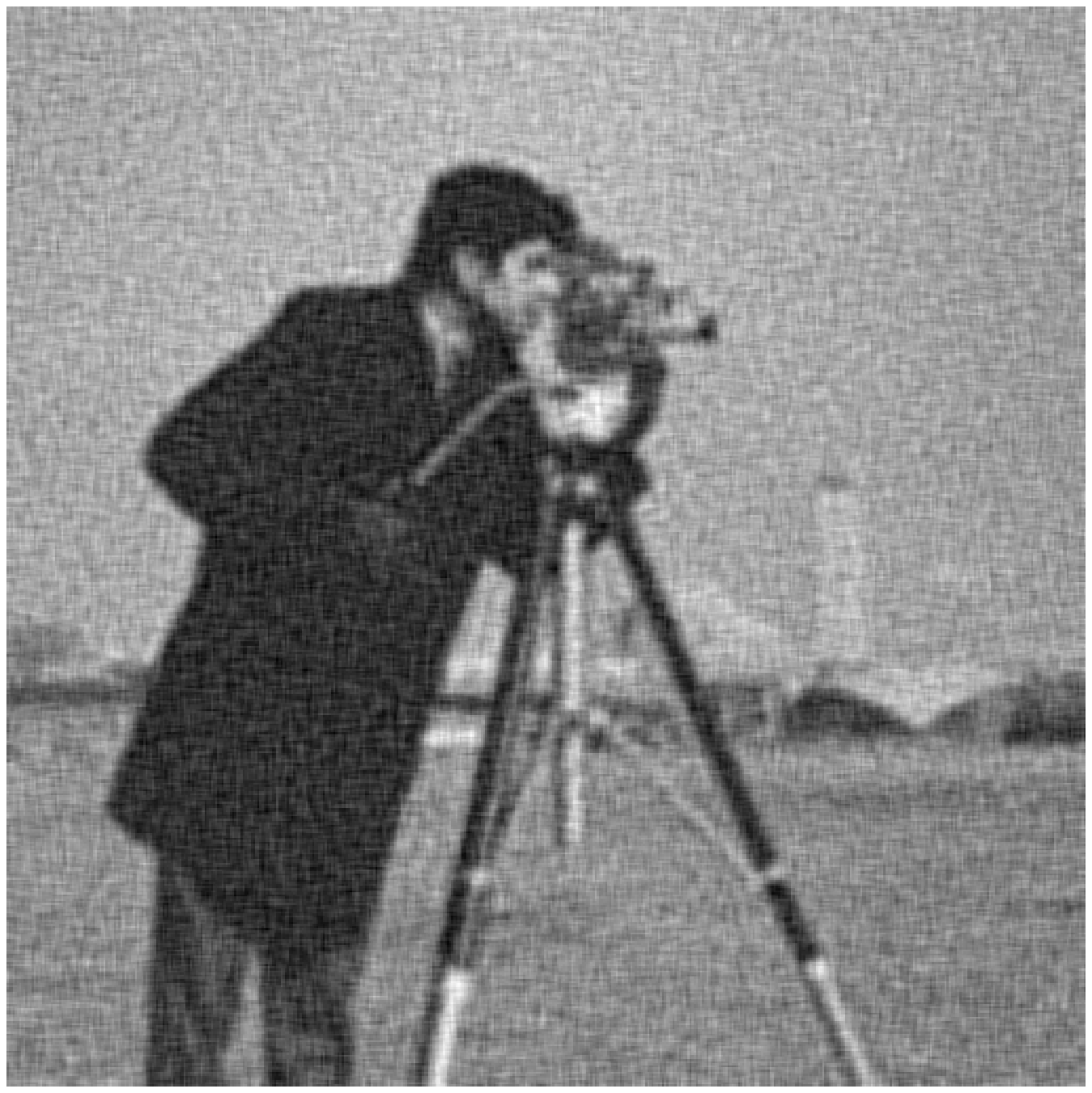}}\resizebox{!}{2
in}{\includegraphics{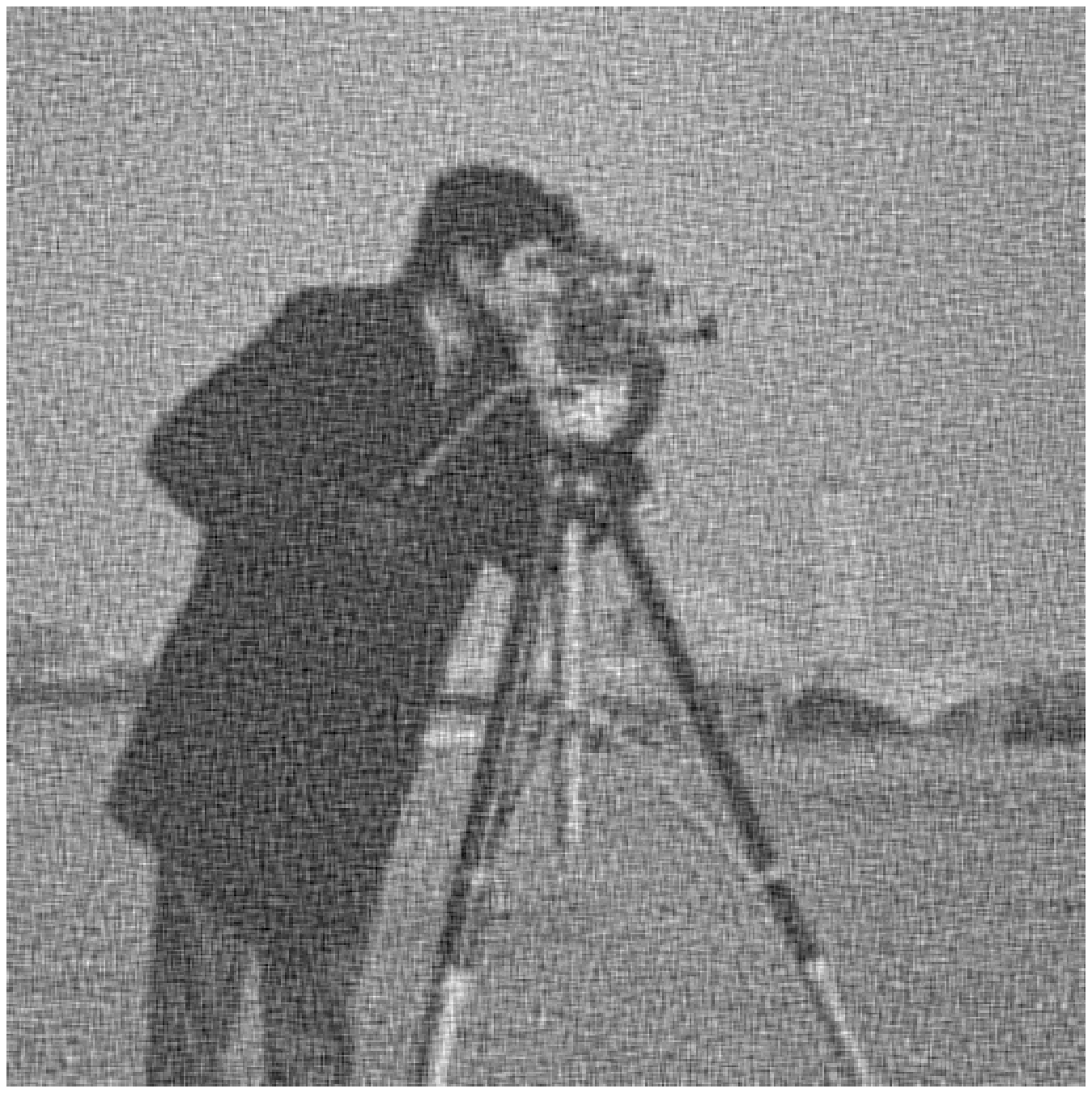}}

 (e) \hspace*{1.9in} (f)

\caption{Deblurring of Cameraman using Tikhonov regularization
with SURE (left) and GCV (right) choices of regularization and
different noise levels: (a), (b) $\sigma=0.01$ (c),(d)
$\sigma=0.05$ (e),(f) $\sigma=0.1$. }
 \label{fig:camera}
\end{center}
\end{figure}

To demonstrate the performance of the proposed regularization
method, we tested it in the context of image deblurring using the
HNO deblurring package for Matlab\footnote{The package is
available at \sloppy
\texttt{http://www2.imm.dtu.dk/\~{}pch/HNO/}.} based on
\cite{HNO06}. We chose several test images, and blurred them using
a Gaussian point-spread function of dimension 9 with standard
deviation 6 using the function \texttt{psfGauss}. We then added
zero-mean, Gaussian white noise with variance $\sigma^2$. In
Figs.~\ref{fig:lena} and \ref{fig:camera} we compare the deblurred
images resulting from using the Tikhonov estimate (\ref{eq:tik})
with $\bbl=\bbi$ where the regularization parameter is chosen
according to our new SURE criterion (left) and the GCV method
(right), for different noise levels.

As can be seen from the figures, our SURE based approach leads to
a substantial performance improvement over the standard GCV
criterion. This can also be seen in Tables~\ref{table:msel} and
\ref{table:msec} in which we report the resulting MSE values.
\begin{table}[h]
\begin{center}
\caption{MSE for Tikhonov Deblurring of Lena} \label{table:msel}
\begin{tabular}{|c|c|c|c|}
\hline \hline & $\sigma=0.01$ & $\sigma=0.05$ & $\sigma=0.1$  \\
\hline \hline
GCV & 0.0022& 0.0077& 0.0133\\
\hline
SURE& 0.0011 & 0.0025 & 0.0042 \\
\hline \hline
 \end{tabular} \end{center}
\end{table}
\begin{table}[h]
\begin{center}
\caption{MSE for Tikhonov Deblurring of Cameraman}
\label{table:msec}
\begin{tabular}{|c|c|c|c|}
\hline \hline & $\sigma=0.01$ & $\sigma=0.05$ & $\sigma=0.1$  \\
\hline \hline
GCV & 0.0033& 0.0121& 0.0221\\
\hline
SURE& 0.0016 & 0.0039 & 0.0064 \\
\hline \hline
 \end{tabular} \end{center}
\end{table}

\subsection{Deconvolution Example}

As another application of the SURE, consider the standard
deconvolution problem in which a signal $\theta[\ell]$ is
convolved by an impulse response $h[\ell]$ and contaminated by
additive white Gaussian noise with variance $\sigma^2$. The
observations $x[\ell]$ can be written in the form of the linear
model (\ref{eq:lg}) where $\bx$ is the vector containing the
observations $x[\ell]$, $\thb$ consists of the input signal
$\theta[\ell]$, and $\bbh$ is a Toeplitz matrix, representing
convolution with the impulse response $h[\ell]$.

To recover $\theta[\ell]$ we may use a penalized LS approach
(\ref{eq:reg}) where we assume that the original signal
$\theta[\ell]$ is smooth.  This can be accounted for by choosing a
penalization of the form $\|\bbl \thb\|_1$ where $\bbl$ represents
a second order derivative operator. The resulting penalized LS
estimate can be determined by solving a quadratic optimization
problem. In our simulations, we used  \texttt{CVX}, a package for
specifying and solving convex programs in Matlab \cite{GB08}.

Since the resulting estimate is non-linear, due to the $\ell_1$
penalization, we cannot apply the GCV equation (\ref{eq:gcv}).
Instead, a popular approach to tune the parameter $\lambda$ is to
use the discrepancy principle in which $\lambda$ is chosen such
that the residual $\|\bx-\bbh\hthb\|^2$ is equal to the noise
level $n\sigma^2$ \cite{GK92,DG95}.

To evaluate the performance of the SURE principle in this context,
we consider an example from the Regularization Tools \cite{H94}
for Matlab.  All the problems in this toolbox are discretized
versions of the Fredholm integral equation of the first kind:
\begin{equation}
\label{eq:fred} g(s)=\int_a^bK(s,t)\theta(t)dt
\end{equation}
where $K(s,t)$ is the kernel and $\theta(t)$ is the solution for a
given $g(s)$. The problem is to estimate $\theta(t)$ from noisy
samples of $g(s)$. Using a midpoint rule with $n$ points,
(\ref{eq:fred}) reduces to an $n \times n$ linear system
$\bx_T=\bbh\thb$. The functions in this toolbox differ in $K(t,s)$
and $\theta(s)$. Below we consider the function  \textsf{heat(n)}
with $n=80$. The output of the function is the matrix $\bbh$ and
the true vector $\thb$ (which represents $\theta(t)$). The
observations are $\bx=\bx_T+\bw$ where $\bw$ is a white Gaussian
noise vector with variance $\sigma^2=1$.

\begin{figure}[h]
\begin{center}
\resizebox{!}{2 in}{\includegraphics{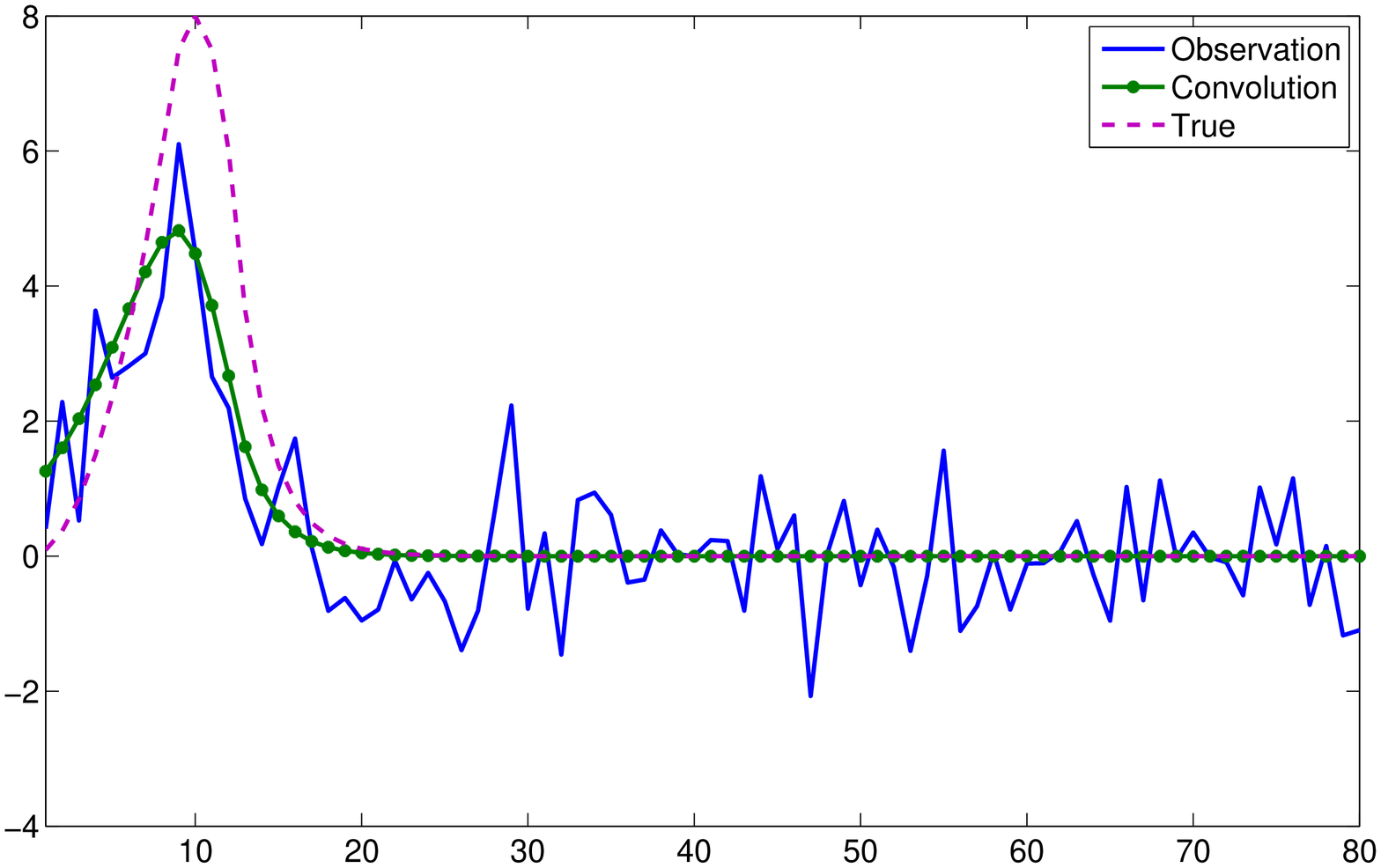}} \caption{The
original signal $\thb$ (dashed), the clean convolved signal (star)
and the observations $\bx$ with $\sigma=1$. }
 \label{fig:heat1}
\end{center}
\end{figure}

\begin{figure}[h]
\begin{center}
\resizebox{!}{2 in}{\includegraphics{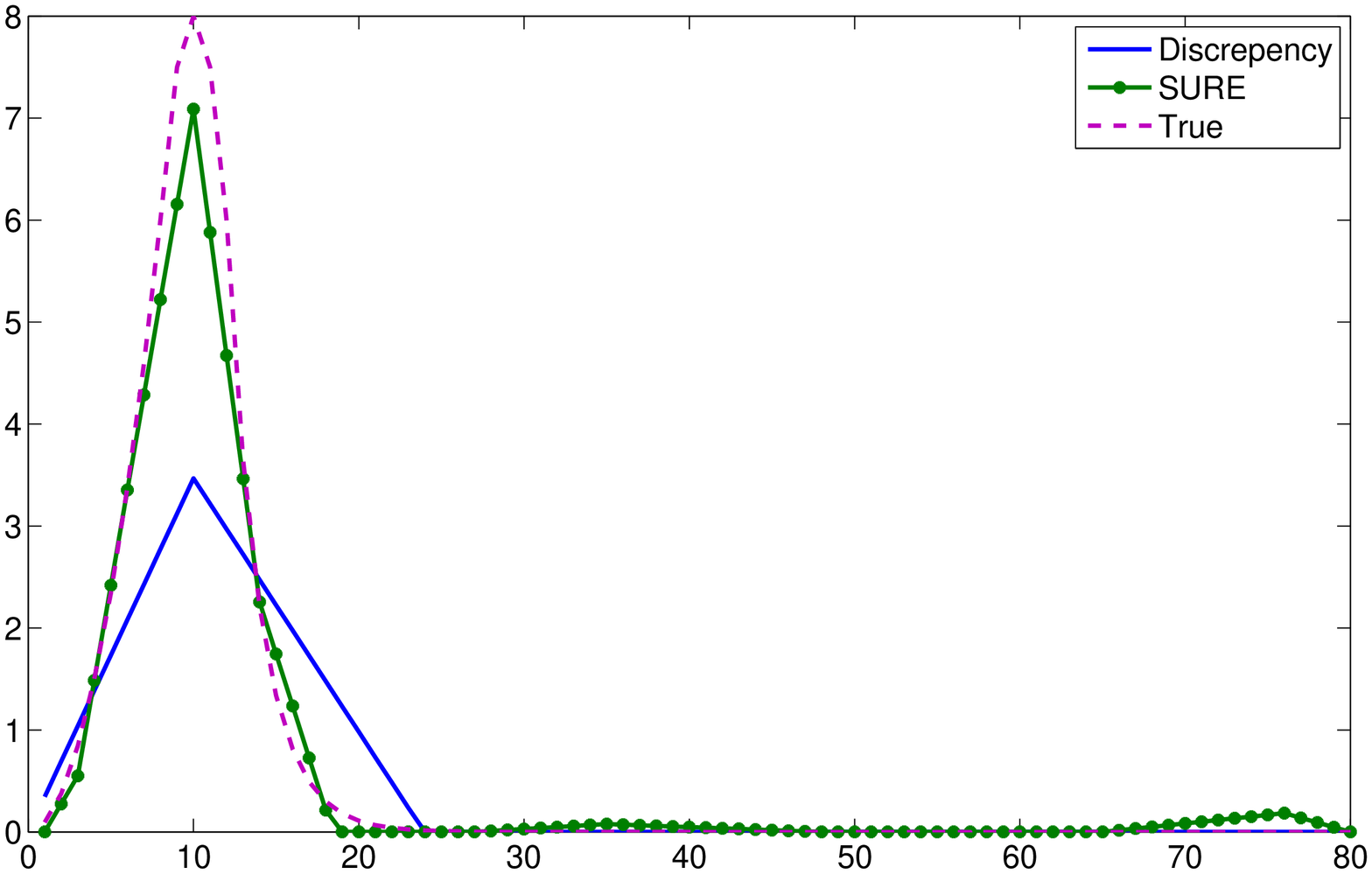}}
\caption{Deconvolution using weighted $\ell_1$ regularization with
the discrepancy principle, SURE (star) and the true signal
(dashed) with $\sigma=1$. }
 \label{fig:heat}
\end{center}
\end{figure}

 In Fig.~\ref{fig:heat1} we plot the original signal along with the
 observations $\bx$, and the clean convolved signal
 $\bx_T=\bbh\thb$.
The original signal along with the estimates using the SURE
principle and  the discrepancy method are plotted in
Fig.~\ref{fig:heat}.
To evaluate the gradient of the estimate, we used the Monte-Carlo
SURE approach proposed in \cite{RBU08}. Evidently, the SURE method
leads to superior performance. The MSE using the SURE approach in
this example is 0.10 while the discrepancy strategy leads to an
MSE of 1.16.

    \section{Regularized SURE Method}
\label{sec:rsure}

A crucial element in guaranteeing success of the SURE method is to
choose a good parameterization of $\bh(\bu)$. However, in many
contexts, such a structure may be hard to find. On the other hand,
letting the SURE criterion select many free parameters can
deteriorate its performance. One way to treat this inherent
tradeoff is by regularization. Thus, instead of minimizing the
SURE objective we suggest minimizing a regularized version:
\begin{equation}
\label{eq:rsure} S(\bh,\lambda)  =  S(\bh)+\lambda r(\bh(\bu))
\end{equation}
where $\lambda$ is a regularization parameter and $ r(\bh(\bu))$
is a regularization function. For example, we may choose
$r(\bv)=\|\bv\|$ where the norm is arbitrary. The parameter
$\lambda$ is determined by applying the conventional SURE
(\ref{eq:spf}) (or (\ref{eq:psure})) to the estimate
$\bh(\bu,\lambda)$ resulting from solving (\ref{eq:rsure}) with
$\lambda$ fixed.

 As an example, consider the iid Gaussian model in which
$\bx=\thb+\bw$ where $\bw$ is a Gaussian noise vector with iid
zero-mean components of variance $\sigma^2$. Assuming that $\thb$
represents the wavelet coefficients of some underlying signal
$\bx$, a popular estimation strategy is wavelet denoising in which
each component of $\bx$ is replaced by a soft or hard-thresholded
version. In particular, in their landmark paper, Donoho and
Johnstone \cite{dj95} developed a soft-threshold wavelet denoising
method in which
\begin{equation}
\label{eq:st} \hat{\theta}_i=\left\{
\begin{array}{ll}
|x_i|-t, & |x_i| \geq t;\\
 0, & |x_i| < t,
\end{array}
 \right.
\end{equation}
where $t$ is a threshold value. They suggest selecting $t$ to
minimize the SURE criterion, and refer to the resulting estimate
as SureShrink (to be more precise, in SureShrink $t$ is determined
by SURE only if it lower than some upper limit). In developing the
SureShrink method, the function $\bh(\bx)$ is restricted to be a
component-wise soft threshold. The motivation for this choice is
that the wavelet coefficients below a certain level tend to be
sparse. It is well known that soft-thresholding can be obtained as
the solution to a LS criterion with an $\ell_1$ penalty:
\begin{equation}
\min\blc \|\bx-\thb\|^2+ \lambda \|\thb\|_1 \brc.
\end{equation}
Thus, in principle we can view the SureShrink approach as a 2-step
procedure: We first determine the estimate that minimizes an
$\ell_1$ penalized LS criterion. We then choose the penalization
factor to minimize SURE.

Instead, we suggest choosing an estimate that directly minimizes
an $\ell_1$ regularized SURE objective, where the only assumption
we make is that the processing is performed component wise. Thus,
$\hat{\theta}_i=\alpha_i x_i$ for some coefficients $\alpha_i(\bx)
\geq 0$. Since $\bu=(1/\sigma^2) x_i$,
$h_i(\bu)=\sigma^2\alpha_iu_i$. With this choice of $\bh(\bu)$,
minimizing (\ref{eq:rsure}) is equivalent to minimizing the
following objective:
\begin{equation}
\label{eq:suret}
S(\mbox{\boldmath{$\alpha$}},\lambda)=\sum_{i=1}^n
\alpha_i^2x_i^2+2\sum_{i=1}^n\alpha_i\bl\sigma^2-x_i^2\br +\lambda
\sum_{i=1}^n |\alpha_i||x_i|.
\end{equation}
The optimal choice of $\alpha_i \geq 0$ is
\begin{equation}
\label{eq:aiopt} \alpha_i=\left[
1-\frac{\sigma^2+\lambda|x_i|}{x_i^2}\right]_+.
\end{equation}
The resulting estimate can be viewed as a soft-thresholding
method, with a particular choice of shrinkage (different than the
standard approach (\ref{eq:st})) when the value of $x_i$ exceeds
the threshold. The precise threshold value is equal to the largest
value $x_i$ for which $\alpha_i=0$ and is given by
\begin{equation}
t=\frac{1}{2}\bl \lambda+\sqrt{\lambda^2+4\sigma^2}\br.
\end{equation}

To choose $\lambda$, we substitute the estimate
$\hat{\theta}_i=\alpha_i(\lambda) x_i$ with $\alpha_i(\lambda)$
given by (\ref{eq:aiopt}) into the SURE criterion (\ref{eq:spg})
with $\bbp=\bbi$, and minimize with respect to $\lambda$. The
value of $\lambda$ can be easily determined numerically.

To demonstrate the advantage of our method over conventional
soft-thresholding we implemented our approach on the examples
taken from \cite{dj95}. Specifically, we used the test functions
Blocks, Bumps, HeaviSine and Doppler defined in \cite{dj95}. The
length of all signals is $2048$ and the noise variance is
$\sigma^2=4$. We used the Daubechies 8 symmetrical wavelet, and
$L=5$ levels are considered. In Table~\ref{table:mse} we report
the empirical MSE values of the original noisy signals, and 3
wavelet denoising schemes: SureShrink which is the method of
\cite{dj95} with the threshold selected using SURE, our proposed
regularized SURE method (RSURE), and OracleShrink which is a
soft-threshold where the threshold value is selected to minimize
the squared-error between the true unknown wavelet coefficient,
and its denoised version. Clearly this approach is only for
comparison and serves as a benchmark on the best possible
performance that can be obtained using any soft threshold.
\begin{table}[h]
\begin{center}
 \caption{MSE for Different Soft Denoising
Schemes} \label{table:mse}
\begin{tabular}{|c|c|c|c|c|}
\hline \hline & Blocks & Bumps & HeaviSine & Doppler \\ \hline
\hline
Original & 4.054& 4.072& 4.153& 3.945\\
\hline
 SureShrink & 0.744& 0.875& 0.205& 0.290\\
\hline
 RSure & 0.694& 0.816& 0.169& 0.273 \\
\hline OracleShrink & 0.690& 0.828& 0.118& 0.283
 \\\hline \hline
 \end{tabular} \end{center}
\end{table}
As can be seen from the table, the regularized SURE method
performs better in all cases than SureShrink. It is also
interesting to see that it sometimes even outperforms OracleShrink
which is based on the true unknown $\thb$. The reason the
performance can be better than the oracle is that the shrinkage
performed in RSURE is different than the conventional soft
threshold.

In Table~\ref{table:mseo} we repeat our experiments where now we
use the estimates resulting from the standard SURE criterion.
Specifically, we consider the positive-part Stein estimate
(\ref{eq:jsm}) referred to as SteinShrink and the estimate
(\ref{eq:jsc}) which we refer to as ScalarShrink.
\begin{table}[h]
\begin{center}
 \caption{MSE for Different Denoising Schemes}
 \label{table:mseo}
\begin{tabular}{|c|c|c|c|c|}
\hline \hline & Blocks & Bumps & HeaviSine & Doppler \\ \hline
\hline
 ScalarShrink & 1.043& 1.362& 0.161& 0.594\\
\hline
SteinShrink& 1.681& 1.730& 1.508& 1.413 \\
\hline \hline
 \end{tabular} \end{center}
\end{table}
Evidently, using the SURE estimate without regularization
deteriorates the performance significantly. Thus, SURE alone is
not generally sufficient to obtain good estimates. However, adding
regularization dramatically improves the behavior without the need
to pre-specify the desired structure.

Finally, in Table~\ref{table:mseh} we repeat the experiments of
Table~\ref{table:mse} to determine the threshold values, but once
the values are found we apply hard-thresholding on the
coefficients.
\begin{table}[h]
\begin{center}
\caption{MSE for Different Hard-Thresholding Schemes}
\label{table:mseh}
\begin{tabular}{|c|c|c|c|c|}
\hline \hline & Blocks & Bumps & HeaviSine & Doppler \\ \hline
\hline
 SureShrink & 1.902& 1.961& 0.988& 0.630\\
\hline
 RSure & 1.560& 1.912& 0.766& 0.700 \\
\hline \hline
 \end{tabular} \end{center}
\end{table}
As can be seen from the table, even though the thresholding
operation is now the same in both methods, RSURE performs
significantly better. Thus, the threshold determined from this
method is superior to that resulting from the SURE criterion
without regularization. Here again the importance of
regularization is demonstrated.

\section{Conclusion}

In this paper, we developed an unbiased estimate of the MSE in
multivariate exponential families by extending the SURE method.
This generalized principle can now be used in exponential
multivariate estimation problems to develop estimators with
improved performance over existing approaches. As an application,
we suggested a new strategy for choosing the regularization
parameter in penalized inverse problems. We demonstrated via
several examples that this method can significantly improve the
MSE over the standard GCV and discrepancy approaches. We also
suggested a regularized SURE criterion for selecting estimators
without the need for pre-specifying their structure. Applying this
objective in the context of wavelet denoising, we proposed a new
type of soft-thresholding which minimizes a penalized estimate of
the MSE. As we demonstrated, this strategy can lead to improved
MSE behavior in comparison with soft and hard thresholding
methods.

The main contribution of this work is in introducing the
generalized SURE criterion and the regularized SURE method and
demonstrating their applicability in several examples. In future
work, we intend to develop these applications in more detail and
further explore the practical use of the proposed design
objectives.

\section{Acknowledgement}

The author would like to thank Zvika Ben-Haim and Michael Elad for
many fruitful discussions.


\begin{thebibliography}{10}

\bibitem{S56}
C.~Stein,
\newblock ``Inadmissibility of the usual estimator for the mean of a
  multivariate normal distribution,''
\newblock in {\em Proc. Third Berkeley Symp. Math. Statist. Prob.} 1956,
  vol.~1, pp. 197--206, University of California Press, Berkeley.

\bibitem{JS61}
W.~James and C.~Stein,
\newblock ``Estimation of quadratic loss,''
\newblock in {\em Proc. Fourth Berkeley Symp. Math. Statist. Prob.} 1961,
  vol.~1, pp. 361--379, University of California Press, Berkeley.

\bibitem{S71}
W.~E. Strawderman,
\newblock ``Proper {B}ayes minimax estimators of multivariate normal mean,''
\newblock {\em Ann. Math. Statist.}, vol. 42, pp. 385--388, 1971.

\bibitem{A73}
K.~Alam,
\newblock ``A family of admissible minimax estimators of the mean of a
  multivariate normal distribution,''
\newblock {\em Ann. Statist.}, vol. 1, pp. 517--525, 1973.

\bibitem{B76}
J.~O. Berger,
\newblock ``Admissible minimax estimation of a multivariate normal mean with
  arbitrary quadratic loss,''
\newblock {\em Ann. Statist.}, vol. 4, no. 1, pp. 223--–226, Jan. 1976.

\bibitem{BE05}
Z.~Ben-Haim and Y.~C. Eldar,
\newblock ``Blind minimax estimators: Improving on least squares estimation,''
\newblock in {\em IEEE Workshop on Statistical signal Processing (SSP'05),
  Bordeaux, France}, July 2005.

\bibitem{BE05j}
Z.~Ben-Haim and Y.~C. Eldar,
\newblock ``Blind minimax estimation,''
\newblock {\em IEEE Trans. Inform. Theory}, vol. 53, pp. 3145--3157, Sep. 2007.

\bibitem{EBN032}
Y.~C. Eldar, A.~Ben-Tal, and A.~Nemirovski,
\newblock ``Robust mean-squared error estimation in the presence of model
  uncertainties,''
\newblock {\em IEEE Trans. Signal Processing}, vol. 53, pp. 168--181, Jan.
  2005.

\bibitem{s73}
C.~M. Stein,
\newblock ``Estimation of the mean of a multivariate distribution,''
\newblock {\em Proc. Prague Symp. Asymptotic Statist.}, pp. 345--381, 1973.

\bibitem{S81}
C.~M. Stein,
\newblock ``Estimation of the mean of a multivariate normal distribution,''
\newblock {\em Ann. Stat.}, vol. 9, no. 6, pp. 1135--1151, Nov. 1981.

\bibitem{dj95}
D.~L. Donoho and I.~M. Johnstone,
\newblock ``Adapting to unknown smoothness via wavelet shrinkage,''
\newblock {\em J. Am. Stat. Assoc.}, vol. 90, no. 432, pp. 1200--1224, Dec.
  1995.

\bibitem{zd98}
X.~P. Zhang and M.~D. Desai,
\newblock ``Adapting denoising based on {SURE} risk,''
\newblock {\em IEEE Signal Process. Lett.}, vol. 5, no. 10, pp. 265--267, 1998.

\bibitem{lbu07}
F.~Luisier, T.~Blu, and M.~Unser,
\newblock ``A new {SURE} approach to image denoising: Interscale orthonormal
  wavelet thresholding,''
\newblock {\em IEEE Trans. Image Process.}, vol. 16, no. 3, pp. 593--606, 2007.

\bibitem{bp05}
A.~Benazza-Benyahia and J.-C. Pesquet,
\newblock ``Building robust wavelet estimators for multicomponent images using
  {S}tein's principle,''
\newblock {\em IEEE Trans. Image Process.}, vol. 14, no. 11, pp. 1814--1830,
  2005.

\bibitem{AH06}
R.~Averkamp and C.~Houdre,
\newblock ``Stein estimation for infinitely divisible laws,''
\newblock {\em ESAIM: Probability and Statistics}, p. 269, 2006.

\bibitem{H78}
H.~M. Hudson,
\newblock ``A natural identity for exponential families with applications in
  multiparameter estimation,''
\newblock {\em Ann. Statist.}, vol. 6, no. 3, pp. 473--484, 1978.

\bibitem{B80}
J.~Berger,
\newblock ``Improving on inadmissible estimators in continuous exponential
  families with applications to simultaneous estimation of gamma scale
  parameters,''
\newblock {\em Ann. Stat.}, vol. 8, no. 3, pp. 545--571, 1980.

\bibitem{H82s}
J.~T. Hwang,
\newblock ``Improving upon standard estimators in discrete exponential families
  with applications to {Poisson} and negative binomial cases,''
\newblock {\em Ann. Statist.}, vol. 10, no. 3, pp. 857--867, 1982.

\bibitem{P36}
E.~Pitman,
\newblock ``Sufficient statistics and intrinsic accuracy,''
\newblock {\em Proc. Camb. phil. Soc.}, vol. 32, pp. 567--579, 1936.

\bibitem{D35}
G.~Darmois,
\newblock ``Sur les lois de probabilites a estimation exhaustive,''
\newblock {\em C.R. Acad. sci. Paris}, vol. 200, pp. 1265--1266, 1935.

\bibitem{K36}
B.~Koopman,
\newblock ``On distribution admitting a sufficient statistic,''
\newblock {\em Trans. Amer. math. Soc.}, vol. 39, pp. 399--409, 1936.

\bibitem{LC98}
E.~L. Lehmann and G.~Casella,
\newblock {\em Theory of Point Estimation},
\newblock New York, NY: Springer-Verlag, Inc., second edition, 1998.

\bibitem{TA77}
A.~N. Tikhonov and V.~Y. Arsenin,
\newblock {\em Solution of Ill-Posed Problems},
\newblock Washington, DC: V.H. Winston, 1977.

\bibitem{R86e}
J.~Rice,
\newblock ``Choice of smoothing parameter in deconvolution problems,''
\newblock {\em Contemporary Math.}, vol. 59, pp. 137--151, 1986.

\bibitem{DG95}
L.~Desbat and D.~Girard,
\newblock ``The {``minimum reconstruction error''} choice of regularization
  parameters: Some effective methods and their application to deconvolution
  problem,''
\newblock {\em SIAM J. Sci. Comput.}, vol. 16, no. 6, pp. 1387--1403, Nov.
  1995.

\bibitem{GK92}
N.~P. Galatsanos and A.~K. Katsaggelos,
\newblock ``Methods for choosing the regularization parameter and estimating
  the noise variance in image restoration and their relation,''
\newblock {\em IEEE Trans. Image Process.}, vol. 1, no. 3, pp. 322--336, 1992.

\bibitem{H93}
P.~C. Hansen,
\newblock ``The use of the {L}-curve in the regularization of discrete
  ill-posed problems,''
\newblock {\em SIAM J. Sci. Stat. Comput.}, vol. 14, pp. 1487--1503, 1993.

\bibitem{HH93}
M.~Hanke and P.~C. Hansen,
\newblock ``Regularization methods for large-scale problems,''
\newblock {\em Surveys Math. Indust.}, vol. 3, no. 4, pp. 253--315, 1993.

\bibitem{B96}
A.~Bj{\"o}rck,
\newblock {\em Numerical Methods for Least-Squares Problems},
\newblock Philadelphia, PA: SIAM, 1996.

\bibitem{MKM99}
R.~Molina, A.~K. Katsaggelos, and J.~Mateos,
\newblock ``Bayesian and regularization methods for hyperparameter estimation
  in image restoration,''
\newblock {\em IEEE Trans. Image Process.}, vol. 8, no. 2, pp. 231--246, 1999.

\bibitem{K05}
W.~C. Karl,
\newblock ``Regularization in image restoration and reconstruction,''
\newblock in {\em Handbook of Image and Video Processing}, A.~Bovik, Ed., pp.
  183–--202. ELSEVIER, 2nd edition, 2005.

\bibitem{ghw79}
G.H. Golub, M.~Heath, and G.~Wahba,
\newblock ``Generalized cross-validation as a method for choosing a good ridge
  parameter,''
\newblock {\em Technometrics}, vol. 21, no. 2, pp. 215--223, May 1979.

\bibitem{K93}
S.~M. Kay,
\newblock {\em Fundamentals of Statistical Signal Processing: Estimation
  Theory},
\newblock Upper Saddle River, NJ: Prentice Hall, Inc., 1993.

\bibitem{LL01}
E.~H. Lieb and M.~Loss,
\newblock {\em Analysis},
\newblock American Mathematical Society, second edition, 2001.

\bibitem{B64}
A.~J. Baranchik,
\newblock ``Multiple regression and estimation of the mean of a multivariate
  normal distribution,''
\newblock Tech. {R}ep.~51, Stanford University, 1964.

\bibitem{HNO06}
P.~C. Hansen, J.~G. Nagy, and D.~P. OLeary,
\newblock {\em Deblurring Images: Matrices, Spectra, and Filtering},
\newblock Philadelphia, PA: SIAM, 2006.

\bibitem{GB08}
M.~Grant and S.~Boyd,
\newblock ``{CVX}: {M}atlab software for disciplined convex programming (web
  page and software),''
\newblock March 2008,
\newblock \texttt{http://stanford.edu/\~{}boyd/cvx}.

\bibitem{H94}
P.~C. Hansen,
\newblock ``Regularization tools, a matlab package for analysis of discrete
  regularization problems,''
\newblock {\em Numerical Algorithms}, vol. 6, pp. 1--35, 1994.

\bibitem{RBU08}
S.~Ramani, T.~Blu, and M.~Unser,
\newblock ``Blind optimization of algorithm parameters for signal denoising by
  {M}onte-{C}arlo {SURE},''
\newblock in {\em Proc. Int. Conf. Acoust., Speech, Signal Processing
  (ICASSP-2008), (Las-Vegas, NV)}, April 2008.

\end{thebibliography}

\newpage
{\bf Yonina C. Eldar} (S'98--M'02--SM'07) Yonina C. Eldar received
the B.Sc. degree in Physics in 1995 and the B.Sc. degree in
Electrical Engineering in 1996 both from Tel-Aviv University
(TAU), Tel-Aviv, Israel, and the Ph.D. degree in Electrical
Engineering and Computer Science in 2001 from the Massachusetts
Institute of Technology (MIT), Cambridge.

From January 2002 to July 2002 she was a Postdoctoral Fellow at
the Digital Signal Processing Group at MIT. She is currently an
Associate Professor in the Department of Electrical Engineering at
the Technion - Israel Institute of Technology, Haifa, Israel. She
is also a Research Affiliate with the Research Laboratory of
Electronics at MIT. Her research interests are in the general
areas of signal processing, statistical signal processing, and
computational biology.

Dr. Eldar was in the program for outstanding students at TAU from
1992 to 1996. In 1998, she held the Rosenblith Fellowship for
study in Electrical Engineering at MIT, and in 2000, she held an
IBM Research Fellowship. From 2002-2005 she was a Horev Fellow of
the Leaders in Science and Technology program at the Technion and
an Alon Fellow. In 2004, she was awarded the Wolf Foundation Krill
Prize for Excellence in Scientific Research, in 2005 the Andre and
Bella Meyer Lectureship, in 2007 the Henry Taub Prize for
Excellence in Research, and in 2008 the Hershel Rich Innovation
Award and Award for Women with Distinguished Contributions. She is
a member of the IEEE Signal Processing Theory and Methods
technical committee, an Associate Editor for the IEEE Transactions
on Signal Processing, the EURASIP Journal of Signal Processing,
and the SIAM Journal on Matrix Analysis and Applications, and on
the Editorial Board of Foundations and Trends in Signal
Processing.

\end{document}